\PassOptionsToPackage{pdfpagelabels=false}{hyperref} 
\documentclass[fleqn,usenatbib]{mnras}

\usepackage{newtxtext,newtxmath}
\usepackage[T1]{fontenc}
\usepackage{ae,aecompl}

\usepackage{graphicx}	% Including figure files
\usepackage{amsmath}	% Advanced maths commands
\usepackage{amssymb}	% Extra maths symbols
\usepackage{hyperref}
\usepackage{color}
\usepackage{multirow}
\usepackage{tikz}
\usepackage{pgfplots, pgfplotstable}
\usepgfplotslibrary{statistics,dateplot}
\usetikzlibrary{patterns}
\pgfplotsset{compat=newest}
\title[Broad-band properties of S2~0109+22]{The broad-band properties of the intermediate synchrotron peaked BL Lac S2~0109+22 from radio to VHE gamma rays}

\author[MAGIC Collaboration]{
\parbox{\textwidth}{MAGIC Collaboration\thanks{Corresponding authors: Fallah Ramazani, V. (\href{mailto:vafara@utu.fi}{vafara@utu.fi}), Hovatta, T. (\href{mailto:talvikki.hovatta@utu.fi}{talvikki.hovatta@utu.fi}), Lindfors, E. (\href{mailto:elilin@utu.fi}{elilin@utu.fi}) and Nilsson, K. (\href{mailto:kani@utu.fi}{kani@utu.fi})}:
S.~Ansoldi$^{1,20}$,
L.~A.~Antonelli$^{2}$,
C.~Arcaro$^{3}$,
D.~Baack$^{4}$,
A.~Babi\'c$^{5}$,
B.~Banerjee$^{6}$,
P.~Bangale$^{7}$,
U.~Barres de Almeida$^{7,8}$,
J.~A.~Barrio$^{9}$,
J.~Becerra Gonz\'alez$^{10}$
% ...
\\
(Affiliations can be found after the references)}
}

\date{Accepted XXX. Received YYY; in original form ZZZ}

\pubyear{2018}
\begin{document}

\label{firstpage}
\pagerange{\pageref{firstpage}--\pageref*{lastpage}}

\maketitle

\begin{abstract}
The MAGIC telescopes observed S2~0109+22 in 2015 July during its flaring activity in high energy gamma rays observed by \textit{Fermi}-LAT. We analyse the MAGIC data to characterise the very high energy (VHE) gamma-ray emission of S2~0109+22, which belongs to the subclass of intermediate synchrotron peak (ISP) BL Lac objects. We study the multi-frequency emission in order to investigate the source classification. Finally, we compare the source long-term behaviour to other VHE gamma-ray emitting (TeV) blazars. We performed a temporal and spectral analysis of the data centred around the MAGIC interval of observation (MJD 57225-57231). Long-term radio and optical data have also been investigated using the discrete correlation function. The redshift of the source is estimated through optical host-galaxy imaging and also using the amount of VHE gamma-ray absorption. The quasi-simultaneous multi-frequency spectral energy distribution (SED) is modelled with the conventional one-zone synchrotron self-Compton (SSC) model. MAGIC observations resulted in the detection of the source at a significance level of $5.3\,\sigma$. The VHE gamma-ray emission of S2~0109+22 is variable on a daily time scale. VHE gamma-ray luminosity of the source is lower than the average of TeV BL Lacs. The optical polarization, and long-term optical/radio behaviour of the source are different from the general population of TeV blazars. All these findings agree with the classification of the source as an ISP BL Lac object. We estimate the source redshift as $z = 0.36 \pm 0.07$. The SSC parameters describing the SED are rather typical for blazars. 
\end{abstract} 

\begin{keywords}
galaxies: active -- galaxies: jets -- gamma rays: galaxies -- BL Lacertae objects: individual: S2~0109+22
\end{keywords}

%%%%%%%%%%%%%%%%%%%%%%%%%%%%%%%%%%%%%%%%%%%%%%%%%%

%%%%%%%%%%%%%%%%% BODY OF PAPER %%%%%%%%%%%%%%%%%%

\section{Introduction}
BL Lac objects dominate the extragalactic very-high-energy (VHE, $E>100$\,GeV) gamma-ray sky. A relativistic jet shoots from the region of the central super-massive black hole, hosted at the center of BL Lac objects, in the line of sight of the observer. Jets are typically characterized by featureless spectra in the optical band, highly polarized radiation in radio and optical, and variable radiation at all frequencies. The jet emission is non-thermal and described as a continuous spectral energy distribution (SED), spanning from radio to VHE gamma-ray frequencies, and featuring two wide peaks. Synchrotron emission by highly relativistic electrons spiralling in the magnetic field of the jet is used to explain the lower frequency peak. Different scenarios within various models are used to explain the high-frequency peak: external Compton \citep{1989ApJ...340..162M, 1994ApJS...90..945D,1994ApJ...421..153S} and  synchrotron self-Compton \citep[SSC,][]{1992MNRAS.258..657C,1992ApJ...397L...5M} as leptonic models, proton synchrotron emission \citep{1996SSRv...75..331M,2000NewA....5..377A, 2001APh....15..121M} and photo-pion production \citep{2014ApJ...797...89A} as hadronic models. Traditionally, in view of their relative simplicity and agreement with the data, single-zone SSC models have been used to describe BL Lac SEDs \citep[e.g.][]{2011ApJ...727..129A,2011ApJ...736..131A}. However, there is growing evidence that these models do not reproduce all the observed features of BL Lac objects \citep[e.g.][]{2014A&A...567A.135A}, and, in some cases, more complicated models should be considered. BL Lac objects are classified according to the peak frequency of their lower energy peak, $\nu_{syn}$ \citep{1994MNRAS.268L..51G}: low synchrotron peaked (LSP; $\nu_{syn}<10^{14}$\,Hz), intermediate synchrotron peaked (ISP; $10^{14}\leq\nu_{syn}<10^{15}$\,Hz), and high synchrotron peaked (HSP; $\nu_{syn}\geq10^{15}$\,Hz) \citep{2010ApJ...716...30A}.

S2~0109+22 (also known as GC~0109+224), at coordinates (J2000) $\text{RA = 01h12m05.8s}$ and $\text{DEC=+22d44m39s}$, was first detected as a compact radio source in the 5\,GHz  Survey  of  the  NRAO  43\,m  dish  of  Green  Bank, West Virginia \citep{1971AJ.....76..980D,1972AJ.....77..265P}. In 1976, it was optically identified as a stellar object of magnitude 15.5 on the Palomar Sky Survey plates, \citet{1977AJ.....82..776O} also measured a strong millimetre emission (1.53\,Jy at 90\,GHz)\footnote{The eleven meter telescope (National Radio Astronomy Observatory) observed the source in 1976.} and defined it as a BL Lac object. Since then it was continuously monitored in radio and optical \citep{2003A&A...400..487C,2008A&A...485...51H,2014AJ....147..143H}. \citet{2003A&A...400..487C,2004MNRAS.348.1379C} performed extensive studies on the radio and optical behaviour and the broad-band SED of this source. It remarkably shows high polarization variability, from 7\% to 30\% \citep{1991A&AS...90..161T,2011ApJS..194...19W}. It is classified as an ISP BL Lac object \citep{1999ApJ...525..127L, 2000A&A...361..480D, 2001MNRAS.325.1109B, 2004MNRAS.348.1379C, 2011ApJ...743..171A} using different approaches and datasets to calculate the location of its synchrotron peak. 

Since the launch of the \textit{Fermi} satellite in 2008, the source has been listed in most of the \textit{Fermi}-LAT catalogues, i.e. 1FGL \citep{2010ApJS..188..405A}; 2FGL \citep{2012ApJS..199...31N}; 1FHL \citep{2013ApJS..209...34A}; and 3FGL \citep{2015ApJS..218...23A}. However, the source is not listed in the catalogue of sources detected $>50\,\text{GeV}$ by the \textit{Fermi}-LAT \citep[2FHL,][]{2016ApJS..222....5A}. The source is variable in the high energy (HE: 100 MeV$<E<$100 GeV) gamma-ray band with the variability index equal to 489 and the maximum monthly flux value of $F_{(0.1-100\,\text{GeV})}=(2.14\pm0.17) \times 10^{-7}$\,ph\,cm$^{-2}$\,s$^{-1}$ which is reported in February 2011 \citep[3FGL,][]{2015ApJS..218...23A}. \citet{2008ApJS..175...97H} reported a redshift value for the source of $z=0.265$, which was disfavoured by \citet{2016MNRAS.458.2836P} using a high signal-to-noise optical spectrum from Gran Telescopio Canarias. Based on this spectrum, $z>0.35$ was measured, assuming the source is hosted by a massive elliptical galaxy typical for this class of sources. VHE gamma-ray observations of this source carried out with MAGIC between 2015 July 22 and 28 (MJD 57225--57231), were triggered when the reported HE gamma-ray daily flux, July 20 (MJD 57223), was about two times higher than the average flux reported in the 3FGL catalogue (private communication with Luigi Pacciani). The MAGIC observations led to the first detection of this source in VHE gamma rays \citep{2015ATel.7844....1M}.

In this paper, we present the multi-frequency observations and data analysis  in Section \ref{sec2}. A long-term behaviour study, the comparison with other VHE gamma-ray emitting (TeV) blazars, and estimations of the source distance are presented in Section \ref{sec3}. Finally, Section \ref{sec4} summarizes our results.
\section{Observations and data analysis}
\label{sec2}
In this section, we introduce the instruments and their respective data analysis procedures.
\subsection{Very high energy gamma rays (MAGIC)}
MAGIC is a system of two Imaging Atmospheric Cherenkov Telescopes (17\,m diameter) located in the Canary Island of La Palma ($28.7^{\circ}$\,N, $17.9^{\circ}$\,W), at the elevation of 2200\,m a.s.l. \citep{2016APh....72...76A}. The use of the stereoscopic technique, combined with large mirror size makes MAGIC one of the most sensitive instruments for VHE gamma-ray astronomy. The corresponding trigger threshold is $\gtrsim 50$\,GeV \citep{2016APh....72...76A}. S2~0109+22 is visible from the MAGIC site at zenith angle below 40$^{\circ}$ between mid-July and February. 

Triggered by increased activity in HE gamma rays, MAGIC observed S2~0109+22 for 9.63\,h in 2015 July within a multi-wavelength blazar monitoring program. The observations were performed during 7 consecutive nights from July 22 to July 28 (MJD 57225--57231) with zenith angle range between 11$^\circ$ and 39$^\circ$. The data have been analysed using the MAGIC Standard Analysis Software \citep[MARS,][]{2009arXiv0907.0943M, 2016APh....72...61A, 2017MNRAS.468.1534A}. Part of the data were affected by clouds, therefore we applied atmospheric transmission correction based on the information obtained with the MAGIC elastic LIDAR \citep{2015EPJWC..8902003F}.

\subsection{High-energy gamma rays (\textit{Fermi}-LAT)}
The Large Area Telescope (LAT) is the primary instrument on-board the \textit{Fermi Gamma-ray Space Telescope}. Based on the pair-conversion technique, it is designed to investigate the gamma-ray sky in the energy band from 30\,MeV to >300\,GeV \citep{2009ApJ...697.1071A}. In its standard operation mode it surveys the sky, covering it fully every 3\,h. 

The data analysed in this paper were selected from a region of interest around S2~0109+22 with a radius of $15^{\circ}$, in a period lasting around three weeks (MJD 57220--57240) roughly centred on the MAGIC detection peak on MJD 57228 (2015 July 25). The data reduction of the events of the Pass8 source class was performed with the ScienceTools software package version v10r0p5\footnote{\url{https://fermi.gsfc.nasa.gov/ssc/data/analysis/software/}} in the energy range 0.1--300\,GeV. To reduce Earth limb contamination a zenith angle cut of 90$^{\circ}$ was applied to the data. The un-binned likelihood fit of the data was performed using the suggested Galactic diffuse-emission model and isotropic component \citep{2016ApJS..223...26A} recommended for Pass8 Source event class\footnote{\url{https://fermi.gsfc.nasa.gov/ssc/data/access/lat/BackgroundModels.html}}.

The normalizations of both diffuse components in the source model were  allowed to freely vary during  the spectral fitting. The source model also includes the sources of the \textit{Fermi}-LAT third source catalogue \citep[3FGL,][]{2015ApJS..218...23A}  within 25$^{\circ}$ of  the source of interest. Spectral indices and fluxes are left to freely vary for sources within 5$^{\circ}$; fluxes are also left to freely vary for sources flagged as 'variable' in the 3FGL catalogue that lie from 5 to 10$^{\circ}$. The spectral parameters of the sources from 10 to 25$^{\circ}$, were instead fixed to their catalogue value.

To construct the  light-curve (LC) with 1-day time bins, only the source  of interest (normalization and spectral index) and  the  diffuse  models (normalization) were left free  to vary, 
while the remaining 3FGL sources were fixed to the values obtained for the three week analysis of the region. An upper-limit is shown when the detection significance was $<3\,\sigma$\footnote{The detection significance for a given source is approximately equal to the square root of the Test Statistic, for a given source.}. The SED was obtained analysing data collected between the 2015 July 22 and 2015 July 28 (MJD 57225--57231), corresponding to the MAGIC observing period.

\subsection{X-ray and UV (\textit{Swift})}
Since 2006, \textit{Neil Gehrels Swift observatory (Swift)} has pointed to the source fifteen times in photon counting mode. Ten of the raw images by the X-ray Telescope \citep[\textit{XRT},][]{2004SPIE.5165..201B} on-board the \textit{Swift} satellite, are qualified for analysis\footnote{\url{https://swift.gsfc.nasa.gov/analysis/threads/gen_thread_attfilter.html}}. The multi-epoch (8) event list for the period from 2015 July 21 (MJD 57224.95) to 2015 August 1 (MJD 57235.86) with a total exposure time of $\sim 6.15$\,h, were downloaded from the publicly available SWIFTXRLOG (\textit{Swift}-XRT Instrument, Log)\footnote{\url{https://heasarc.gsfc.nasa.gov/W3Browse/swift/swiftxrlog.html}}. These observations have an average integration time of 2.8\,ks each. They were processed using the procedure described by \citet{2017A&A...608A..68F}, assuming fixed equivalent Galactic hydrogen column density $n_H = 4.24 \times 10^{20}\,\rm cm^{-2}$ reported by \citet{2005A&A...440..775K}. Additionally, \textit{Swift} observed this source two more times in 2006. We analysed those two additional event lists to get a broader view of the source's X-ray properties.

The Ultraviolet/Optical Telescope (\textit{UVOT}, $[4.9-16.6] \times 10^{5}$\,GHz) on-board the \textit{Swift} satellite \citep{2008MNRAS.383..627P}, observed the source 15 times during the MAGIC campaign, out of which eight were simultaneous to the \textit{XRT} data taking\footnote{The difference between the number of data points measured by \textit{UVOT} and \textit{XRT} is due to the usage of \textit{XRT} window timing mode, multiple \textit{UVOT} snapshots during \textit{XRT} exposure, and bad quality of \textit{XRT} raw images.}. An iterative data calibration procedure \citep{2010A&A...524A..43R} was used to calculate the Galactic extinction\footnote{Calculated based on the value obtained from \citet{2011ApJ...737..103S}}, the effective frequency, and the flux conversion factor for each filter.

\subsection{Optical}

\subsubsection{Light-curve (KVA, KAIT, and Catalina)}
S2~0109+22 was added to the Tuorla blazar monitoring program\footnote{\url{http://users.utu.fi/kani/1m}} when HE activity was reported in 2015 July. The monitoring observations were performed in optical R-band using a 35\,cm Celestron telescope coupled to the KVA (Kunglinga Vetenskapsakademi) telescope located at La Palma. Data analysis was performed using a semi-automatic pipeline for differential photometry assuming the comparison star magnitudes in \citet{2003A&A...400..487C}. The magnitudes were corrected for Galactic extinction using values from \citet{2011ApJ...737..103S}.

In order to study the long-term optical behaviour of S2~0109+22, its optical LC is retrieved from the publicly available online database of 76-cm Katzman Automatic Imaging Telescope (KAIT) at Lick Observatory\footnote{\url{http://herculesii.astro.berkeley.edu/kait/agn}}. The LC from KAIT is produced through a pipeline that utilizes aperture photometry and performs brightness calibrations using USNO B1.0 catalogue stars in the source field. The long-term optical LC is extended back to 2005 by including available online data from the Catalina Real-Time Transient Survey \citep{2009ApJ...696..870D}. KAIT and Catalina data are obtained from unfiltered observations, whose effective color is close to the R-band \citep{2003PASP..115..844L}.

\subsubsection{Host galaxy imaging (NOT)}
%\textcolor{red}{Under construction by Kari}
\label{HGIMAGE}
To investigate the host galaxy of S2~0109+22, we obtained a deep I-band image at the Nordic Optical Telescope (NOT) on 2015 November 11. In total, 26 exposures, each 150 seconds long, were obtained using the ALFOSC\footnote{\label{ALFOSC}\url{http://www.not.iac.es/instruments/alfosc}} instrument.  After subtracting the bias, flat-fielding and fringe map correction, the images were registered using stars in the field and summed. The resulting image has a total exposure time of 3900 seconds with $FWHM \cong 1.14\arcsec$. The comparison stars in \citet{2003A&A...400..487C} were used to calibrate the field.

\subsubsection{Polarization (NOT)}
Polarization observations were carried out using the ALFOSC instrument in the standard linear polarization set-up (lambda/2 retarder followed by calcite) in optical \textit{R}-band. Weekly observations were performed from November 2015 to  September 2017 within three observing seasons. In order to determine the zero point of the position angle, polarization standards were observed on a monthly basis. The instrumental polarization was measured observing zero-polarization standard stars, and was found to be negligible. Most of the observations were conducted under good sky condition (seeing $\sim 1\arcsec$). 

Using aperture (radius of $1.5\arcsec$) photometry, the sky-subtracted target counts were measured for ordinary and extraordinary beams. By using the intensity ratios of two beams and standard formulae in \citet{2007ASPC..364..495L}, we calculated normalized Stokes parameters, polarization fraction, and position angle for each observation. Systematic uncertainties are included in our error estimation.

\subsection{Radio (OVRO and Mets\"ahovi)}

S2~0109+22 was observed at 15\,GHz as part of a high-cadence gamma-ray blazar monitoring program using the Owens Valley Radio Observatory (OVRO) 40~m telescope \citep{2011ApJS..194...29R}. The observations are calibrated by using a temperature-stable diode noise source to remove receiver gain drifts, and the flux density scale is derived from observations of 3C~286 assuming the value of 3.44\,Jy at 15.0\,GHz \citep{1977A&A....61...99B}. The systematic uncertainty of about 5\% in the flux density scale is not included in the error bars. Complete details of the reduction and calibration procedure are found in \citet{2011ApJS..194...29R}.

The Mets{\"a}hovi radio telescope, operating at 37\,GHz, has been observing the source for two decades. We selected radio data obtained after mid-2005 for the long-term study of the source. The instrument and data reduction procedures are described by \citet{1998A&AS..132..305T}.

\section{Results}
\label{sec3}

\subsection{Very high energy gamma rays}
\label{sec3-1}
\begin{figure}
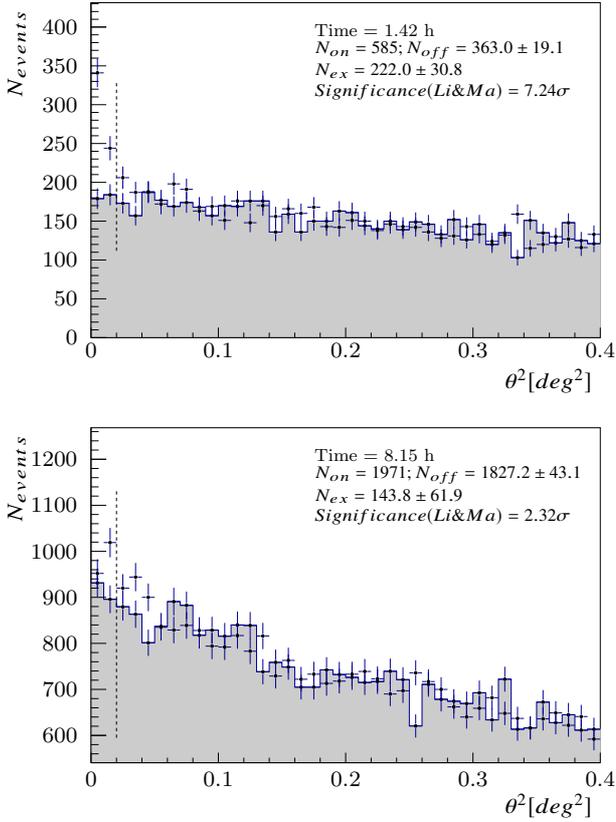

\resizebox{\linewidth}{!}{\input{odie_20150725.tex}}
\resizebox{\linewidth}{!}{\input{odie_nonflare.tex}}
\caption{\label{theta2} $\theta^2$ distribution of the S2~0109+22 events, signal (blue) and background (shadowed grey), for the 1.4\,h of MAGIC observations during the flare night (top) and for all other observations 8.2\,h (bottom). The vertical dashed line indicates the defined signal region.}
\end{figure}

The VHE gamma-ray signal from the source is estimated after applying energy dependent selection cuts to the signal. Residual background of the observation is measured around a control region \citep{2017MNRAS.468.1534A}. The distribution of the events is shown in Figure \ref{theta2}. In total, there was an excess of ($365.8\pm 69.1$) events in the signal region $\theta^2 < 0.02\,\deg^2$, where $\theta^2$ is the squared angular distance between the reconstructed source position of the events and the nominal position of the expected source. The data taken during MJD 57228 (2015 July 25) contribute  $\geq 61\%$ of excess events of the whole sample of data. The source was detected at a significance level of $7.24\,\sigma$ during MJD 57228 (Fig. \ref{theta2}). 

The LC of the VHE gamma-ray integral flux above 100\,GeV ($F_{>100\,\text{GeV}}$) is shown in Figure \ref{fig_lc} with the details presented in Table \ref{tab_tev_flux}. The constant flux hypothesis is disfavoured with $\chi^2/\text{d.o.f.} = 14.5/4$ ($P_{value}=0.005$). The peak flux, detected on MJD 57228 (hereafter flare night), is twice the average flux over the whole period of observation, $F_{>100\,\text{GeV, ave}}=~(4.7 \pm 1.2) \times 10^{-11}$\,ph\,cm$^{-2}$\,s$^{-1}$. %The lowest observed flux, during 1 week of MAGIC campaign, is detected on MJD 57225. 
Real correlation analysis for such a short period around the flare night is beyond the reach with the available data sample shown in Figure \ref{fig_lc}. However, there seems to be an increased flux in X-rays, optical and UV bands around the flare night, which suggests that emission in these bands could originate from a single region.

\input{lcmaker.tex}

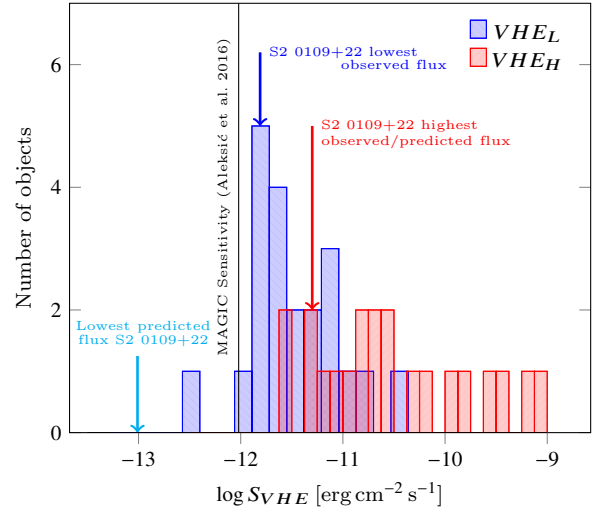
\begin{figure}
\begin{tikzpicture}
\begin{axis}[ybar, ymin=0, ymax=7,
             xlabel={$\log S_{VHE}$\,[erg\,cm$^{-2}$\,s$^{-1}$]},
             ylabel={Number of objects},
             xtick pos=left,
             legend style={at={(0.975,0.975)},anchor=north east,
                           fill=none,draw=none}
             ]

\addplot +[fill opacity=.2,color=blue,
           postaction={pattern color=blue,pattern=north west lines},
           hist={bins=25, data min=-13.25, data max=-9}   
           ] 
           table [y index=6,col sep=comma] {fdist.csv};

\addplot +[fill opacity=.20,color=red,
           postaction={pattern color=red,pattern=north east lines},
           hist={bins=28, data min=-12.5, data max=-9}   
           ] 
           table [y index=5,col sep=comma] {fdist.csv};
\legend{$VHE_{L}$,$VHE_{H}$}

\draw [<-,blue,line width=0.35mm] (axis cs:-11.81,5) -- (axis cs:-11.81,6.2) node [right] {{\tiny S2~0109+22 lowest}};

\node at (axis cs:-10.5,5.8) [above,blue] {{\tiny observed flux}};

\draw [<-,cyan,line width=0.35mm] (axis cs:-13.01,0) -- (axis cs:-13.01,1.25); 
\node [cyan] at (axis cs:-12.95,1.5) {{\tiny  flux S2~0109+22}};
\node at (axis cs:-12.95,1.5) [above,cyan] {{\tiny Lowest predicted}};

\draw [<-,red,line width=0.35mm] (axis cs:-11.30,2) -- (axis cs:-11.30,5) node [right] {{\tiny S2~0109+22 highest}};

\node at (axis cs:-11.30,4.75) [right,red,line width=0.35mm] {{\tiny observed/predicted flux}};

\draw [black] (axis cs:-12.02,0) -- (axis cs:-12.02,7) node [rotate=90, pos=0.55, above] {{\tiny MAGIC Sensitivity (Aleksi{\'c} et al. 2016)}};

\draw [black] (axis cs:-9.,0) -- (axis cs:-13.5,0);
\end{axis}
\end{tikzpicture}
\caption{\label{histo2}Different states of the observed and predicted VHE $\gamma$-ray flux (>200\,GeV) of S2~0109+22 compared to the  distribution of the variable TeV BL Lac sample  reported in \citet{2017A&A...608A..68F}. This sample contains BL Lac  objects with at least two flux measurements in  VHE gamma rays.}
\end{figure}

We compare the integral flux ($F_{>200\,\text{GeV}}$) of S2~0109+22 to that of other TeV BL Lac objects (21 sources) with variable flux in VHE gamma rays presented in the most recent population study by \citet{2017A&A...608A..68F}, who studied a time independent correlation between several lower-energy bands and VHE gamma-ray luminosity, and predicted the VHE gamma-ray flux for 182 non-TeV BL Lac objects. The comparison is shown in Figure \ref{histo2}. Both the lowest and the largest observed flux of S2~0109+22 during the MAGIC campaign are among the faintest of the population. High and low state VHE gamma-ray predicted energy flux (>200\,GeV) in \citet{2017A&A...608A..68F} are $(4.5\pm1.9)\times 10^{-12}$ and $(9.8\pm2.1)\times 10^{-14}$\,erg\,cm$^{-2}$\,s$^{-1}$, respectively. The largest observed flux over the same energy range, $F_{>200\,\text{GeV}}^{\text{high~obs}}=(4.6\pm1.5)\times 10^{-12}$\,erg\,cm$^{-2}$\,s$^{-1}$, is in good agreement with the predicted flux. The lowest observed flux of this source is $F_{(>200\,\text{GeV})}^{\text{low obs}}=(1.5\pm0.7)\times 10^{-12}$\,erg\,cm$^{-2}$\,s$^{-1}$. The observed VHE gamma-ray flux of the source is fainter than the sample of variable TeV BL Lacs.

Figure \ref{vhe_sed} shows the spectrum of S2~0109+22 in the VHE gamma rays. We assume a simple power-law model, 
\begin{equation}
\frac{dN}{dE}=F_0 (\frac{E}{E_{\text{dec}}})^{-\Gamma},
\end{equation}
where $E_{\text{dec}}$ and $F_0$ are the decorrelation energy and differential flux at $E_{\text{dec}}$, and $\Gamma$ is the spectral photon index. The spectral parameters are obtained via forward-folding using Poissonian maximum likelihood procedure described by \citet{2017MNRAS.472.2956A}. In order to calculate the intrinsic spectral parameters, the same estimation procedure is used by assuming $z=0.35$ (see Sect. \ref{sec3-5}) and Extragalactic Background Light (EBL) absorption model described by \citet{2011MNRAS.410.2556D}. The spectral parameters are summarized in Table \ref{tab_tev_spec} for the flare night and the average spectrum. The fitted model statistics are calculated in the energy range of 65--370\,GeV and 65--250\,GeV for average and flare night spectra, where MAGIC detected the source.

\subsection{High-energy gamma rays}
\label{sec3-2}
We have found that there is no significant HE gamma-ray spectral and flux variability on a daily basis during the investigated period (MJD 57220--57240). These results are shown in Figure \ref{fig_lc} (Panels b and c). The HE gamma-ray constant fit flux is $F_{(0.1-300\,\text{GeV})} = (1.4 \pm 0.4) \times 10^{-7}$\,ph\,cm$^{-2}$\,s$^{-1}$, which is $\sim\,2$ times higher than the average flux reported in the 3FGL catalogue \citep{2015ApJS..218...23A} for this source. 

To model the HE gamma-ray spectrum of S2~0109+22, a power-law function which uses integrated flux as a free parameter\footnote{\url{https://fermi.gsfc.nasa.gov/ssc/data/analysis/scitools/source_models.html\#PowerLaw2}} is used. 
\begin{equation}
\frac{dN}{dE}=\frac{N(\Gamma +1)E^{\Gamma}}{E^{\Gamma +1}_{\text{max}}-E^{\Gamma +1}_{\text{min}}}
\end{equation}
where $\Gamma$ is the photon index, $E_{\text{min}}=100\,\text{MeV}$, $E_{\text{max}}=300\,\text{GeV}$, and N is the integral flux between $E_{\text{min}}$ and $E_{\text{max}}$.

\begin{figure}
\begin{tikzpicture}
\pgfplotsset{width=0.43\textwidth}
\usepgfplotslibrary{fillbetween}

\begin{loglogaxis}[name = magic,
             every axis plot post/.append style = {error bars/.cd, 
                                                   y dir=both,
                                                   y explicit,
                                                   x dir=both,
                                                   x explicit
                                                   },
             legend style={at={(0.025,0.975)},anchor=north west,
                           fill=none,draw=none,font = \tiny},
             ytick pos=left,
             xmin = .081,   
             xmax = 1100,
             ymin = 5e-13,
             ymax = 1.6e-9,
             scaled x ticks = false,
             legend image post style={sharp plot,-},
             ylabel = {$\log E^2.\frac{dF}{dE}$\,[TeV\,cm$^{-2}$\,s$^{-1}$]},
             xlabel = {$\log E~[\rm GeV]$},
             xticklabels = {,-1,0,1,2,3},
             yticklabels = {,-12,-11,-10,-9}
             ]
\addplot[only marks,color =black,]
         table
         [    
           x expr=\thisrowno{0}, 
           y expr=\thisrowno{1},
           y error expr=\thisrowno{2}
         ] {jul25_sed};
\addplot[black] [domain=65:300, samples=10,unbounded coords=jump]{(pow((x/119.43),-3.69))*(11.66*(10^-10))*(pow(x,2))*(pow(0.001,2))};

\addplot[only marks, mark= o,color =gray]
         table
         [ 
           x expr=\thisrowno{0}, 
           y expr=\thisrowno{3},
           y error expr=\thisrowno{4}
         ] {jul25_sed};  
\addplot[gray] [domain=65:300, samples=10,unbounded coords=jump]{(pow((x/119.43),-3.07))*(15.63*(10^-10))*(pow(x,2))*(pow(0.001,2))};

\addplot[only marks,mark=square*, color=blue]
         table
         [ 
           x expr=\thisrowno{0}, 
           y expr=\thisrowno{1},
           y error expr=\thisrowno{2}
         ] {stack_sed};
\addplot[blue] [domain=65:500, samples=10,unbounded coords=jump]{(pow((x/137.13),-3.45))*(2.48*(10^-10))*(pow(x,2))*(pow(0.001,2))};

\addplot[color =cyan,only marks,mark=square]
         table
         [ 
           x expr=\thisrowno{0},
           y expr=\thisrowno{3},
           y error expr=\thisrowno{4}
         ] {stack_sed};         
\addplot[cyan] [domain=65:500, samples=10,unbounded coords=jump]{(pow((x/130.95),-2.92))*(4.15*(10^-10))*(pow(x,2))*(pow(0.001,2))};

%\addplot+[color =red,only marks,mark=triangle*]
%         table
%         [ 
%           x expr=\thisrowno{0}, 
%           y expr=\thisrowno{1},
%           y error expr=\thisrowno{2}
%         ] {fermi_spec.txt};         

%\addplot[red] [domain=0.12:30, samples=10,unbounded coords=jump]{((pow(x*3.8,2))*((1.4e-10)*pow(x,-1.81))/(pow(350,0.81)-pow(0.1,0.81)))};

\addplot[red, mark=none, nodes near coords, point meta=explicit symbolic] 
         table
         [ 
           x expr=\thisrowno{0}, 
           y expr=\thisrowno{3},
           col sep = comma
         ] {fermi_bow.csv}; 
\addplot[name path=A, red!30, mark=none, nodes near coords, 
         point meta=explicit symbolic] 
         table
         [ 
           x expr=\thisrowno{0}, 
           y expr=\thisrowno{5},
           col sep = comma
         ] {fermi_bow.csv}; 
\addplot[name path=B, red!30, mark=none, nodes near coords, 
         point meta=explicit symbolic] 
         table
         [ 
           x expr=\thisrowno{0}, 
           y expr=\thisrowno{6},
           col sep = comma
         ] {fermi_bow.csv};
\addplot[red!30] fill between[of=A and B];         

\addplot[color =red,only marks,mark=triangle*]
         table
         [ x expr=\thisrowno{0},
           y expr=\thisrowno{1},
           y error expr=\thisrowno{2},
           x error plus expr={\thisrowno{4}- \thisrowno{0}},
           x error minus expr={\thisrowno{0}- \thisrowno{3}}
         ] {fermi_spec2.txt}; 
         
\draw[->, color=red](axis cs:56.1,7.94338125E-011)--(axis cs:56.1,4.94338125E-011);  
\draw[->, color=red](axis cs:232,3.390118125E-010)--(axis cs:232,2.090118125E-010);

\legend{MAGIC Flare observed,,MAGIC Flare intrinsic,,MAGIC stacked observed,,MAGIC stacked intrinsic,,,,,,Fermi intrinsic}
\end{loglogaxis}
\begin{loglogaxis}[xmin = 0.081,    
                   xmax = 1100,
                   hide x axis,
                   ymin = 8e-13,
                   ymax = 2.56e-9,
                   axis y line*=right,
                   yticklabels = {,-12,-11,-10,-9},
                   ylabel = {$\log \nu F_{\nu}$\,[erg\,cm$^{-2}$\,s$^{-1}$]}
                   ]
\end{loglogaxis}
\end{tikzpicture}
\caption{\label{vhe_sed}The observed (filled symbols) and intrinsic spectrum (open symbols) of the source obtained from MAGIC data for the flare night (MJD 57228, circles) and for all observations (MJD 57225--57231, squares) together with the HE gamma-ray spectrum obtained from \textit{Fermi}-LAT data (MJD 57225--57232, triangles). The VHE gamma-ray spectra are corrected for the EBL absorption effect using the \citet{2011MNRAS.410.2556D} model.}

\end{figure}
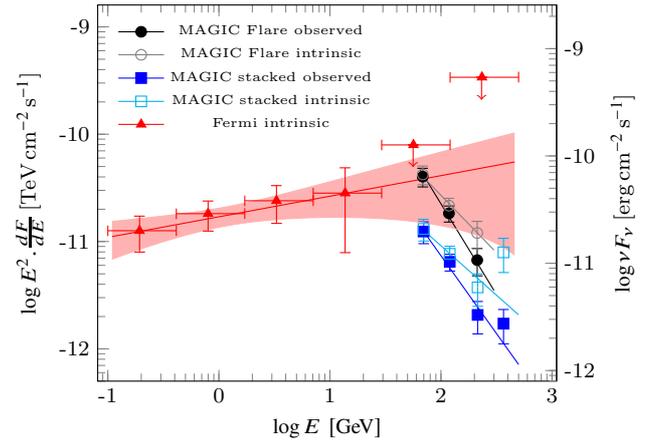

We analysed the source in the period MJD 57225--57232, modelling its spectrum with a simple power-law. The likelihood fit obtained a Test Statistic of $TS= 111$. The resulting power law index of the fitted model is $\Gamma=1.81 \pm 0.14$. The spectral index of the investigated period is within the error bars of the one reported in 3FGL. In Figure \ref{vhe_sed}, we show the flux values in six logarithmically spaced bins from 100\,MeV to 300\,GeV. Upper-limits are shown when the detection significances are lower than $3\,\sigma$. 

%\documentclass{article}
%\usepackage[margin=0.2cm]{geometry}
%\usepackage{tikz}
%\usepackage{pgfplots}

%\begin{document}
%\pdfpagewidth 21.cm   %set one column width for A&A
%\pdfpageheight 9.cm  %set the total hieght of figure. 

%\thispagestyle{empty}  
%\textbf{•}

%\end{document}

\subsection{X-rays}
\label{xray3}
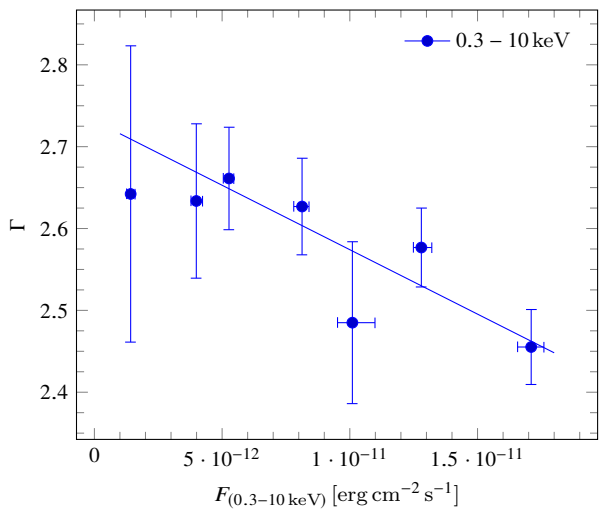
\begin{figure}
\begin{tikzpicture}
\pgfplotsset{
             scaled ticks = false ,
             minor y tick num = 3,
             minor x tick num = 4,
             legend style={at={(0.975,0.975)}, anchor=north east,
                           fill=none, draw=none},
             every axis plot post/.append style = {error bars/.cd, 
                                                   y dir=both,
                                                   y explicit,
                                                   x dir=both,
                                                   x explicit
                                                   },
             unbounded coords=jump
             }

\begin{axis}[name = xray, 
             ylabel = {$\Gamma$},
             xlabel = {$F_{(0.3-10\,\text{keV})}$\,[erg\,cm$^{-2}$\,s$^{-1}$]},
             legend image post style = {sharp plot,-}]
\addplot+ [only marks] table [y = Xrt-index, 
                y error  = Xrt-index-err, 
                x = f0.3-10, 
                x error plus = up-err-f0.3-10, 
                x error minus = low-err-f0.3-10,
                col sep = comma] {database2.csv}; 
\addlegendentry{$0.3-10\,\text{keV}$}
\addplot[blue] [domain=1e-12:1.8e-11, samples=10,unbounded coords=jump]{((1/-6.35007880E-11)*x)+(1.73460898E-10/6.35007880E-11)};

\end{axis}

%\begin{axis}[xlabel = {$\Gamma$},ylabel= {$F[ph/cm^2/s]$},at=(xray.below south west),anchor=north west]
%\addplot+ [only marks] table [x = Index, 
%                x error  = Err_Index,
%                y = flux, 
%                y error  = Err_Flux, 
%                col sep = comma] {fermi.csv}; 
%\addlegendentry{$\Gamma_{(0.1-300 GeV)}$}
%\end{axis}

\end{tikzpicture}
\caption{\label{spec_flux}X-ray spectral index vs. flux during the MAGIC campaign. The blue line shows the best fitted linear model.}
\end{figure}

The results of our X-ray analysis are shown in Figure \ref{fig_lc} with the details available in Table \ref{tab_xray}. The X-ray flux peaks three nights before the VHE gamma-ray peak. The X-ray spectrum is usually soft (Photon index, $\Gamma_X \geq 2.4$). The constant flux hypothesis is rejected with $>10\,\sigma$ level of confidence. However, only a hint of brighter-harder trend with $2\,\sigma$ level of confidence is present in our data sample. The trend between X-ray spectral index and flux ($F_{0.3-10\,\text{keV}}$) can be described by a linear model (Fig. \ref{spec_flux}) with the test statistics of $\chi ^2/\text{d.o.f.}= 2.97/5$, corresponding to Pearson correlation coefficient of 0.76. Moreover, we tried to fit a log-parabola model to the data obtained on MJD 57228. It reveals that the power law model with an index $\Gamma_X = 2.58 \pm 0.05$ ($\chi ^2/\text{d.o.f.}= 48.2/50$) can describe the spectrum better. The X-ray flux ($F_{0.3-10\,\text{keV}}$) on the flare night was $>6$ times higher than the flux from 2006 observations.

\subsection{Long-term behaviour}

Recently two studies of optical and radio behaviour of TeV blazars have been published. \citet{2016A&A...593A..98L} studied the long-term optical and radio behaviour of 32 VHE gamma-ray blazars using data from the OVRO and Tuorla blazar monitoring programs. They found correlated flares in half of the sources, and correlated long-term trends in 13 sources. \citet{2016A&A...596A..78H} performed a first statistical study of the optical polarization variability of TeV blazars, and found that they are not different from the control sample of non-TeV blazars. S2~0109+22 was not part of those studies. In order to compare its optical and radio behaviour with the sample of VHE gamma-ray blazars, we have performed the same analysis of the long-term optical and radio data and optical-polarization data as done in \citet{2016A&A...593A..98L} and \citet{2016A&A...596A..78H}. 

Moreover, the long-term correlation studies between radio/optical and gamma-ray bands were already performed by \citet{2014MNRAS.445..437M} and \citet{2014ApJ...797..137C} using similar radio and optical datasets as those presented in this analysis. Therefore, we only attempt to study the long-term radio-optical cross-correlation behaviour of the source together with its optical polarization behaviour.

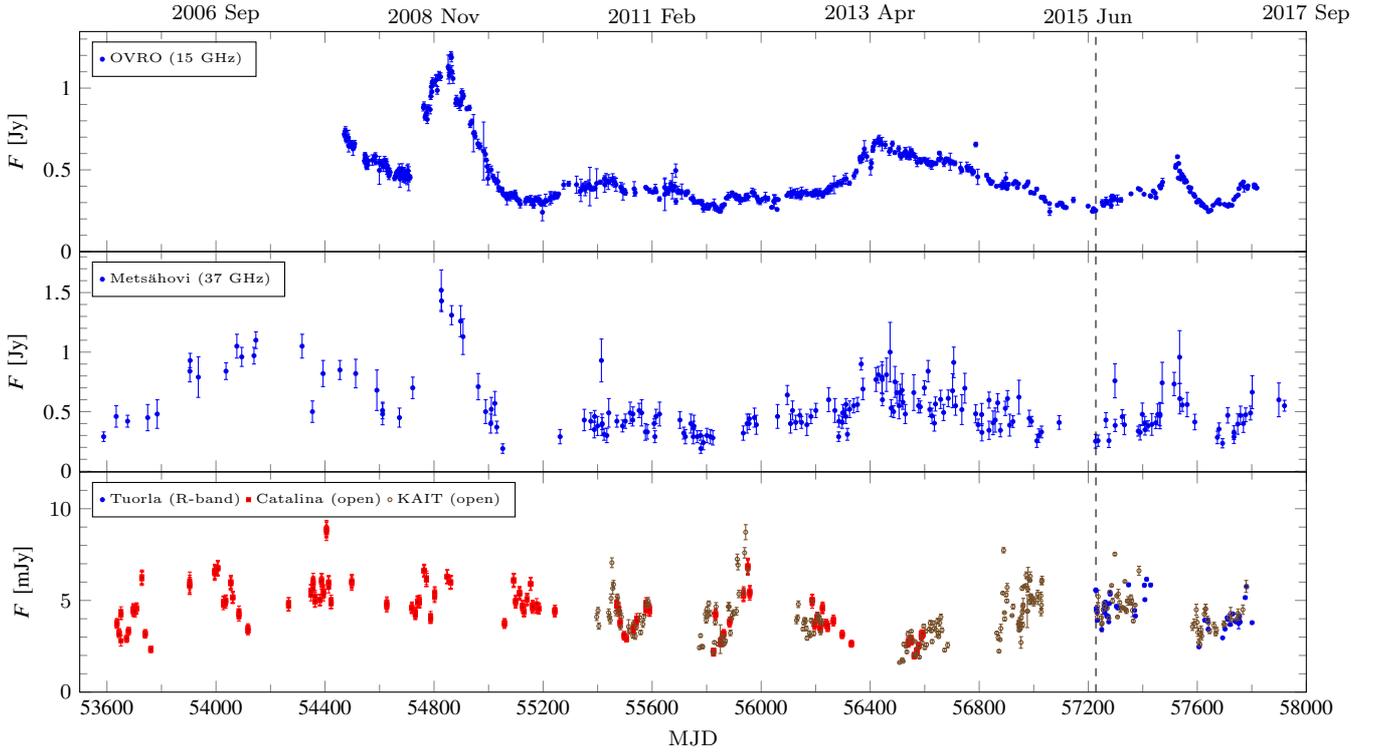
\begin{figure*}
\begin{tikzpicture}
\pgfplotsset{scaled ticks = false ,
             minor y tick num = 4,
             minor x tick num = 1,
             xtick distance=400,
             %ylabel style = {font=\tiny},
             %x tick label style = {font=\tiny},
             legend style={at={(0.01,0.955)},legend columns=3,
                           anchor=north west,fill=none,
                           font = \tiny},
             every axis plot post/.append style = {
                                                   mark size=0.75, 
                                                   only marks,
                                                   error bars/.cd, 
                                                   y dir=both,
                                                   y explicit,
                                                   x dir=both,
                                                   x explicit
                                                   },
             unbounded coords=jump,
             xticklabel style={
                     /pgf/number format/fixed,       
                     /pgf/number format/precision=0,
                     /pgf/number format/1000 sep={}
                     },
             every node near coord/.style={font=\tiny},
             width = \textwidth,
             height = 4.5cm ,
             xmin=53500,
             xmax=58000
             }
%%%%%% OVRO Flux with symetric error bars %%%%%
    
\begin{axis}[name = ovro,
             xticklabel pos=right,
             xticklabels={,,,2006 Sep,,2008 Nov,,2011 Feb,
             ,2013 Apr,,2015 Jun,,2017 Sep},
             ymin=-0.0001,
             ylabel = {$F~[\rm Jy]$}                  
             ]

\addplot+[nodes near coords, point meta=explicit symbolic] 
          table [x = mjd,
                y = flux, 
                y error = flux_err] {ovro.txt}; 
\addlegendentry{OVRO (15 GHz)}
\draw [dashed] (axis cs:57228,0) -- (axis cs:57228,1.9);
\end{axis}

%%%%%% Metsa Flux with symetric error bars %%%%%

\begin{axis}[name = mets,
             xticklabels = {,} ,
             at=(ovro.below south west),
             yshift = 0.05cm,
             anchor=north west,yshift = 0.15cm,
             ylabel = {$F~[\rm Jy]$}           
             ]
\addplot+[nodes near coords, point meta=explicit symbolic] 
          table [
                x = mjd, y = Flux, 
                y error  = Error] {Mesahovi_S20109+22.txt}; 
\addlegendentry{Mets{\"a}hovi (37 GHz)}
\draw [dashed] (axis cs:57228,0) -- (axis cs:57228,1.9);
\end{axis}

%%%%%% Tuorla Flux with symmetric error bars %%%%%

\begin{axis}[name = optical,xlabel = MJD,
             at=(mets.below south west),
             anchor=north west,ymin=-0.0,ymax=0.012,
             yshift = 0.2cm,
             ylabel = {$F~[\rm mJy]$},
             yticklabels = {,0,5,10}               
             ]
\addplot table [x expr= {\thisrow{jd}-2400000.5}, y = flux, 
                y error = error] {KVA_S2_0109+22.txt}; 
\addlegendentry{Tuorla (R-band)}

\addplot+[nodes near coords, point meta=explicit symbolic]
         table [x =mjd, y = mjy, 
                y error = mjyerr,col sep = comma] {cata.csv}; 
\addlegendentry{Catalina (open)}

\addplot+[nodes near coords, point meta=explicit symbolic, mark=o]
         table [x expr= {\thisrow{jd}-2400000.5}, y = mjy, 
                y error = mjyerr,col sep = comma] {kait.csv}; 
\addlegendentry{KAIT (open)}

%\addplot table [x = MJD, y = JY, 
%                y error = JY_Error] {Frank_C.txt}; 
%\addlegendentry{Frankfurt (C?-band)}

%\addplot table [x = MJD, y = JY, 
%                y error = JY_Error] {Frank_V.txt}; 
%\addlegendentry{Frankfurt (V-band)}
\draw [dashed] (axis cs:57228,0) -- (axis cs:57228,0.012);
\end{axis}
\end{tikzpicture}
\caption{\label{fig_longterm}Long-term radio and optical LC of S2~0109+22. \textit{Top:} Radio flux density at 15\,GHz (OVRO). \textit{Middle:} Radio flux density at 37\,GHz (Mets\"ahovi). \textit{Bottom:} Optical flux density in R-band (KVA) and with open filter (KAIT and Catalina). The vertical dashed line indicates the flare night (MJD 57228).}
\end{figure*}

\subsubsection{Radio-Optical cross-correlation analysis}
\label{3-4-1}

Figure \ref{fig_longterm} illustrates the long-term optical and radio data of S2~0109+22. The coverage is of 12 years in the optical band (R-band and open filters) and at 37\,GHz, and 10 years at 15\,GHz.

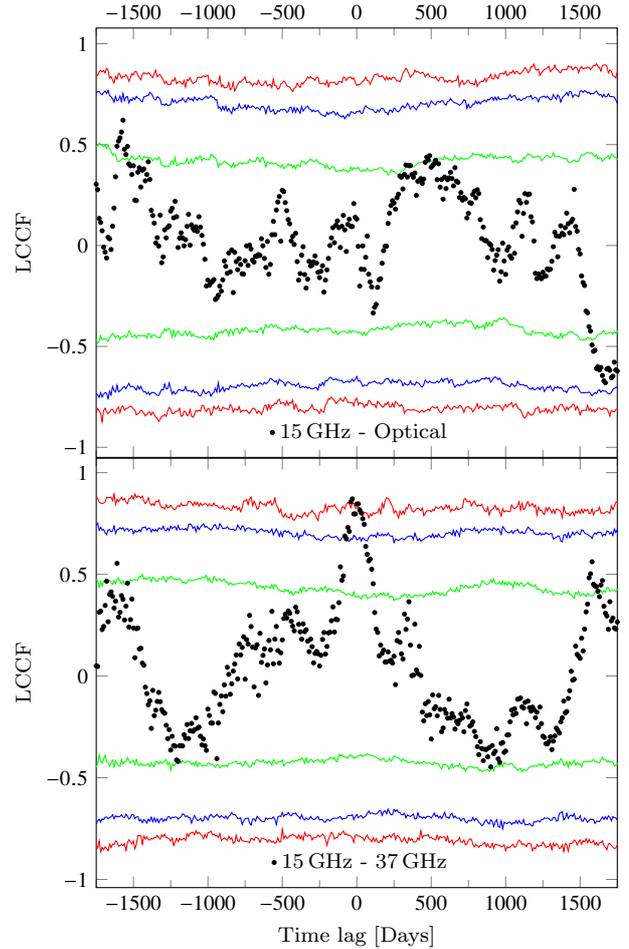
\begin{figure}
\begin{tikzpicture}
\pgfplotsset{
             compat = newest ,
             scaled ticks = false,
             ticklabel style={/pgf/number format/fixed,
                              /pgf/number format/1000 sep={}},
             every node near coord/.style={font=\tiny},
             legend style={at={(0.7,0.105)}, draw=none, fill =none},
             }   
\begin{axis}[name = lccf,
             xmin = -1750,    
             xmax = 1750, 
             ylabel = {LCCF},
             minor x tick num = 1,
             xticklabel pos=right, 
             ]
\addplot[only marks, color=black, mark size=0.75]
         table [ x = col1, y = col3, col sep = comma] {crosscorr_ovro_optical_cut.csv}; 
\addplot[color= green, mark size=0.5]
         table [ x = del, y = col1, col sep = comma] {crosscorr_significance_ovro_optical_cut.csv}; 
\addplot[color= green, mark size=0.5]
         table [ x = del, y = col2, col sep = comma] {crosscorr_significance_ovro_optical_cut.csv};
\addplot[color= blue, mark size=0.5]
         table [ x = del, y = col3, col sep = comma] {crosscorr_significance_ovro_optical_cut.csv}; 
\addplot[color= blue, mark size=0.5]
         table [ x = del, y = col4, col sep = comma] {crosscorr_significance_ovro_optical_cut.csv}; 
\addplot[color= red, mark size=0.5]
         table [ x = del, y = col5, col sep = comma] {crosscorr_significance_ovro_optical_cut.csv}; 
\addplot[color= red, mark size=0.5]
         table [ x = del, y = col6, col sep = comma] {crosscorr_significance_ovro_optical_cut.csv};

\addlegendentry{15\,GHz - Optical}
\end{axis}

\begin{axis}[name = lccfrr,
             at=(lccf.below south west),
             anchor=north west,
             yshift=0.05cm,
             xmin = -1750,    
             xmax = 1750, 
             ylabel = {LCCF},
             xlabel = {Time lag [Days]},
             minor x tick num = 1,
             %yticklabels = {,0,0.5,1.0,1.5},
             %ytick = {-14,-13,-12,-11,-10,-9},
             %xtick = {10,12,14,16,18,20,22,24,26}
             %extra y tick labels = {0.25,0.75, 1.25}
             ]
\addplot[only marks, color=black, mark size=0.75]
         table [ x = col1, y = col3, col sep = comma] {crosscorr_ovro_metsahovi.csv}; 
\addplot[color= green, mark size=0.5]
         table [ x = del, y = col1, col sep = comma] {crosscorr_significance_ovro_metsahovi.csv}; 
\addplot[color= green, mark size=0.5]
         table [ x = del, y = col2, col sep = comma] {crosscorr_significance_ovro_metsahovi.csv};
\addplot[color= blue, mark size=0.5]
         table [ x = del, y = col3, col sep = comma] {crosscorr_significance_ovro_metsahovi.csv}; 
\addplot[color= blue, mark size=0.5]
         table [ x = del, y = col4, col sep = comma] {crosscorr_significance_ovro_metsahovi.csv}; 
\addplot[color= red, mark size=0.5]
         table [ x = del, y = col5, col sep = comma] {crosscorr_significance_ovro_metsahovi.csv}; 
\addplot[color= red, mark size=0.5]
         table [ x = del, y = col6, col sep = comma] {crosscorr_significance_ovro_metsahovi.csv};

\addlegendentry{15\,GHz - 37\,GHz}
\end{axis}
\end{tikzpicture}
\caption{\label{lccf}\textit{Top:} The results of the DCF study between optical (R-band) and radio (15 GHz). \textit{Bottom:} The results of the DCF study between radio bands (15 and 37 GHz); We show $1\,\sigma$, $2\,\sigma$ and $3\,\sigma$ significance limits (green, blue, and red lines, respectively). Positive significant lags show that the flare at 15 GHz is leading the other bands.}
\end{figure}

Following \citet{2016A&A...593A..98L}, we calculated the cross-correlation function between the optical and 15\,GHz LCs using the Discrete Correlation Function \citep[DCF;][]{1988ApJ...333..646E} with local normalization \citep[LCCF;][]{1999PASP..111.1347W}. We use temporal binning of 10 days and require that each LCCF bin has at least 10 elements. Following \citet{2014MNRAS.445..437M}, the significance of the correlation is estimated using simulated LC. In the simulations, we used a power spectral density index of -1.8 for the radio LC (Max-Moerbeck et al. in prep.), which is slightly smaller than the values between -1.4 and -1.7 reported in \citet{2004MNRAS.348.1379C} for the (8 to 37\,GHz) radio LCs. For the optical, we used a power spectral density index of -1.5 (Nilsson et al. in prep.). While there are several peaks (features) in the LCCF, shown in Figure \ref{lccf}, none of them reach the $2\,\sigma$ significance level. We also calculated the cross-correlation functions between the optical-37\,GHz and 37-15\,GHz. The only significant correlation is between 37-15\,GHz, with significance $>3\sigma$. The peak is rather broad from -40 to +30 days (Fig. \ref{lccf}) and is consistent with zero lag. Typically, for evolving synchrotron self-absorbed components \citep[e.g.][]{1994ApJ...437...91S,2014MNRAS.441.1899F}, one would expect the higher frequency to lead the lower frequency variations, which is consistent with our finding. However, as stated the peak is rather wide and also consistent with zero time lag. These results may indicate co-spatiality.

The optical-radio correlations of this source have been previously studied by \citet{2002A&A...394...17H} and \citet{2004MNRAS.348.1379C}. Both works found several weak peaks in the correlations with lags 190, 400 days \citep{2002A&A...394...17H}, and 190, 789 and 879 days \citep{2004MNRAS.348.1379C}. In Figure \ref{lccf}, there is a single `feature' covering all these lags, peaking at $\sim$500 days. This feature is not significant and in general the results of our calculation agree with those by \citet{2002A&A...394...17H} and \citet{2004MNRAS.348.1379C}.

We also searched for common long-term trends from the optical and radio data by fitting linear trends to these LCs. No long-term trends were found at these wavelengths.

We then compared the results of the correlation and trend analyses to the results obtained for other TeV blazars in \citet{2016A&A...593A..98L}. The sources in which no connection between flaring behaviour nor long term behaviour were found were a minority in that sample and were either very weak sources, or bright sources with clear outbursts like S2~0109+22. These other bright sources in the \citet{2016A&A...593A..98L} sample were S5~0716+714, ON~325 and W~Com and it was suggested that as there were several $2\,\sigma$ peaks in their correlation function, there might be several time-scales involved, blurring the correlation. However, for S2~0109+22 we do not find any correlation peaks above $2\,\sigma$. This result may indicate that a major fraction of the optical flux in this source is not originating from the same emission region as the radio, or that the radio-optical correlation is more complex than can be probed by the simple cross-correlation function used in this paper. 

\subsubsection{Optical Polarization}

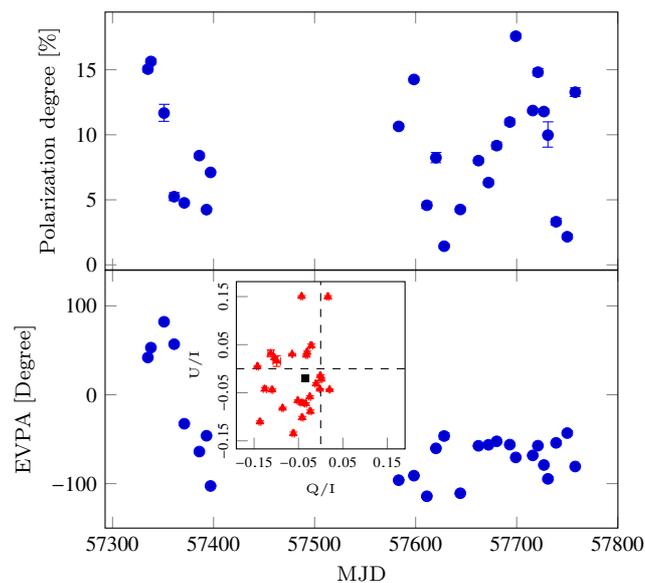
\begin{figure}
\begin{tikzpicture}
\pgfplotsset{compat=newest,set layers=standard,
             scaled ticks = false ,
             every axis plot post/.append style = {only marks,
                                                   error bars/.cd, 
                                                   y dir=both,
                                                   y explicit,
                                                   x dir=both,
                                                   x explicit
                                                   },
             unbounded coords=jump,
             every node near coord/.style={font=\tiny}
             }
\begin{axis}[name = pd,
             anchor=north west,
             xticklabels = {,}, 
             ylabel = {Polarization degree [\%]}, 
             width = 0.47\textwidth,
             height = 5cm 
             ]
\addplot table [x expr= {\thisrow{jd}-2400000.5}, 
                y = pd, 
                y error plus expr = {\thisrow{perrup}-\thisrow{pd}}, 
                y error minus expr = {\thisrow{pd}-\thisrow{perrlo}}]
                {pol}; 

\end{axis}
\begin{axis}[name = PA,
             at=(pd.below south west),
             anchor=north west,ymax=140,
             yshift = 0.2cm,ymin=-150,
             xlabel = MJD ,%tick pos=left,
             ylabel = {EVPA [Degree]},
             width = 0.47\textwidth,
             height = 5cm,
             xticklabel style={/pgf/number format/fixed,
                               /pgf/number format/precision=0,
                               /pgf/number format/1000 sep={}},              
             ]
\addplot table [x expr= {\thisrow{jd}-2400000.5}, y = pa, 
                y error = paer] {pol};
\end{axis}

\begin{axis}[name = small,
             at=(PA.north east),
             anchor=north east,
             mark size=1.5, 
             tick pos=left,
             xtick={-.15,-.05,.05,.15},
             label style={font=\tiny},
             ticklabel style={/pgf/number format/fixed, font=\tiny},              
             ytick={-.15,-.05,.05,.15},
             yticklabel style={rotate=90},
             xmin=-0.19,
             xmax=0.19,
             width = 3.8cm,
             height = 3.8cm ,
             xshift = -2.8cm,
             yshift = 0.05cm,
             xlabel = {Q/I},
             ylabel = {U/I}              
             ]
\addplot[mark =triangle*,color=red] table [x expr= {\thisrow{px}*.01},
                y expr= {\thisrow{py}*.01}, 
                x error expr= {\thisrow{pxerr}*.01},
                y error expr= {\thisrow{pyerr}*.01}
                ] {pol};
\node[fill,inner sep=1.5pt] at (axis cs:-0.034,-0.0190) {};;      
\draw [dashed] (axis cs:0,-.19) -- (axis cs:0,0.19);
\draw [dashed] (axis cs:-.19,0) -- (axis cs:0.19,0);
\end{axis}
\end{tikzpicture}
%\vspace{-5.4cm}
\caption{\label{pollc}\textit{Top panel:} Degree of polarization in the optical (R-band) obtained with the Nordic Optical telescope. \textit{Bottom panel:} Same for the polarization angle. The inset shows the source polarization measurements in $Q/I$-$U/I$-plane. The black square in the $Q/I$-$U/I$-plane is the mass center of weighted average of the $Q/I$ and $U/I$ values.}
\end{figure}

The optical emission in active galaxies is dominated by synchrotron emission of the their jet, which is intrinsically highly polarized. In an optically thin jet with uniform magnetic field, the polarization fraction can be up to 70\% \citep[e.g.,][]{1970ranp.book.....P}. The more typically observed levels of fractional polarization reach a few tens of percent at maximum \citep[e.g.,][]{1980ARA&A..18..321A,2016MNRAS.463.3365A}, which have been taken as evidence for disordered magnetic fields. The linearly polarized emission is described using the Stokes parameters \textit{I} (for total intensity), and \textit{Q} and \textit{U} (for linear polarization). Using the Stokes parameters, the polarization fraction and the electric vector position angle (EVPA) can be defined as $m=(\sqrt{Q^2+U^2})/I$ and EVPA$=1/2\tan^{-1}(U/Q)$. The polarization fraction and EVPA for S2~0109+22 are shown in Figure \ref{pollc}.

We estimate the long-term polarization variability of S2~0109+22 by using the methods described in \citet{2016A&A...596A..78H} where the optical polarization of a sample of TeV and non-TeV-detected BL Lac objects was studied. We calculate the intrinsic mean polarization fraction and its modulation index (standard deviation of the polarization fraction over the mean), by assuming that the polarization fraction follows a Beta distribution, which is confined to values between 0 and 1, similarly as the polarization fraction. A single polarization observation is assumed to follow a Ricean distribution, so that our probability density function is obtained by convolving the Beta and Ricean distributions as follows,

\vspace{-.25cm}
\begin{equation}
{\rm PDF}\left(p;\alpha, \beta\right)=\frac{p^{\alpha -1} \left( 1-p\right)^{\beta -1}}{B\left(\alpha, \beta\right)},
\end{equation}
where $p$ is the polarization fraction and $\alpha$ and $\beta$ determine
the shape of the Beta distribution $B\left(\alpha, \beta\right)$. If
the parameters $a, \beta$ of this distribution are known, the mean
polarization fraction and the intrinsic modulation index are then given by 

\vspace{-.25cm}
\begin{equation}
p_\text{int}=\frac{\alpha}{\alpha + \beta}
\end{equation}
and
\vspace{-.25cm}
\begin{equation}
m_\text{int}=\frac{\sqrt{Var}}{p_\text{int}}=\frac{\sqrt{\frac{\alpha\beta}{\left(\alpha + \beta\right)^2 \left(\alpha + \beta
       +1 \right)}}}{\frac{\alpha}{\alpha + \beta}}, 
\end{equation}
where $Var$ is the variance of the distribution. Details of the method are described in Appendix A of \citet{2016MNRAS.457.2252B}. The intrinsic mean polarization fraction of S2~0109+22 is $0.090^{+0.010}_{-0.008}$, which is higher than the sample mean values of $0.054\pm0.008$ and $0.079\pm0.009$ obtained for the TeV and non-TeV BL Lac objects in \citet{2016A&A...596A..78H}. Similarly, the intrinsic modulation index of the polarization fraction $0.54^{+0.08}_{-0.06}$ is higher than the sample mean values for the TeV ($0.29\pm0.03$) and non-TeV ($0.38\pm0.04$) sources. 

The polarization angle variability can be quantified by calculating the derivative of the polarization angle variations. First we account for the n$\pi$ ambiguity of the polarization angle by requiring that each subsequent point is within 90$^\circ$ from the previous observation. We obtain a median derivative of 2.4 degrees per day, which translates to 3.3 degrees per day in the source frame when multiplied by $(1+z)$ ($z=0.35$, see Sect. \ref{sec3-5}). Comparing this to the histograms in Figure~4 of \citet{2016A&A...596A..78H} shows how S2~0109+22 varies more rapidly in polarization angle than the average TeV (mean $1.11\pm0.29$ deg./day) and non-TeV (mean $1.66\pm0.45$ deg./day) sources. This is also seen when we examine the polarization variations in the $Q/I-U/I$-plane (see the inset in Fig. \ref{pollc} for the $Q/I-U/I$ plot). As described in \citet{2016A&A...596A..78H} a tightly clustered distribution of the points in the $Q/I-U/I$-plane is an indication of a preferred polarization angle. For S2~0109+22 the weighted average of the $Q/I$ and $U/I$ values places the mass center at a distance of 0.039 from the origin, which is smaller than the mean value of $0.050\pm0.008$ for the TeV sources in \citet{2016A&A...596A..78H}. However, the spread in the points, quantified as the distance of each point from the mass center, is 0.077, which is much higher than the mean values ($0.021\pm0.003$ for TeV and $0.041\pm0.005$ for non-TeV sources) in \citet{2016A&A...596A..78H}. In fact, there is only one non-TeV source with a value higher than we obtain for S2~0109+22.

These results are in good agreement with previous studies \citep[e.g.][]{1991A&AS...90..161T} and indicate that the optical polarization of S2~0109+22 is more variable both in fractional polarization and position angle than other high-energy BL Lac objects, and that there does not seem to be a preferred polarization angle in the source, at least over our monitoring period. This is not unexpected based on the analysis of \citet{2016A&A...596A..78H} which showed that the polarization variability depends more on the position of the synchrotron peak rather than the detection of TeV emission \citep[see also][]{2015A&A...578A..68C}. As shown in Figure \ref{SED}, in the ISP-type S2~0109+22 the optical emission probes the peak of the synchrotron component, where the variability is expected to be higher \citep[see also][]{2016MNRAS.463.3365A}. Comparing the obtained intrinsic mean polarization fraction to the values presented by \citet{2016MNRAS.463.3365A}, this source seems to be a rather typical ISP-type object. The maximum polarization fraction is over 15\%, which is high, but not uncommon for ISP sources, as shown in \citet{2016A&A...596A..78H} where about 30\% of the ISP objects reach fractional polarization values as high as or higher than 15\%. This indicates that the magnetic field order must be fairly high in the emission region.

\subsection{Redshift estimation}\label{sec3-5}
The lack of emission lines in the optical spectrum of BL Lacs objects makes the determination of the redshift of these sources particularly challenging. An estimation on the distance can be obtained from basic assumption on the host galaxy luminosity \citep[e.g.][]{2003A&A...400...95N}. Alternatively, an upper limit on the distance can instead be estimated by studying the deformation induced by the EBL on the VHE gamma-ray spectrum.

\subsubsection{Host galaxy}
\label{sec3-5-1}
We use the deep I-band image (see \ref{HGIMAGE}) to search for the host galaxy emission. Two-dimensional surface brightness models were fitted to the light distribution of S2~0109+22 in order to study its host galaxy. Prior to the fitting, the background level was measured and subtracted, removing also a small tilt in the background. Two models were considered: 1) a point source (jet) model and 2) a point source + elliptical galaxy model. Both models had three free parameters, point source x-y position plus flux in the first model, and point source flux, host galaxy flux and host galaxy effective radius in the second model. The first model was used to fix the position of the nucleus, i.e. the second model was fit using the position from the first model to fix the point source and the host galaxy into the same position. Moreover, the ellipticity of the host galaxy was fixed to $\varepsilon = 0$ and the Sersi\'c index to $n = 4$. Both models were convolved with the PSF, determined from two nearby stars, located at 61\arcsec and 84\arcsec away from, and with similar peak intensity to S2~0109+22. The fit was performed using pixels within 10.5\arcsec of the center of S2~0109+22.

We used a Metropolis sampler \citep[e.g.][]{2017arXiv170404629M} to map a posteriori distribution in three-dimensional parameter space. We employed 10 independent walkers, each completing 30~000 iteration steps and with flat priors. The walkers were initially distributed randomly over a fairly wide range of values, but they all quickly converged towards the same area in the parameter space corresponding to the maximum likelihood.  The calculation of likelihood assumed that the pixel values had an uncertainty consisting of four components, each normally distributed: 1) Photon noise, 2) readout noise, 3) error in background determination and 4) error in the PSF model. The background uncertainty was determined by measuring the background around the source in 10 rectangular regions. For the PSF error, we subtracted the PSF from a star close to S2~0109+22 and examined the residuals. The residuals were the strongest near the center of the star, where they amounted to 2\% of the local signal.

Figure \ref{hostplot} shows the marginalized posterior distributions of the two host galaxy parameters: the host galaxy flux and effective radius. The parameters are correlated and in addition both correlate strongly with the point source flux. The best-fit (mode of the posteriors) parameters of model no.2 correspond to AGN flux = (6.651$\pm$0.003) mJy, host galaxy flux (0.149$\pm$0.003) mJy and effective radius (1.40$\pm$0.04)\arcsec. The host galaxy flux in the \textit{I}-band optical is I = 18.05 mag.

If we make the assumption that the host galaxy is a passively evolving early type galaxy with absolute magnitude $M_R = -22.8$ \citep{2005ApJ...635..173S} with $R - I$ = 0.7 and using $A_I$ = 0.057 for the Galactic absorption \citep{2011ApJ...737..103S}, then we obtain $z = 0.36 \pm 0.07$. This value and its error are a result of 1000 trials where we first drew $M_R$ from a Gaussian distribution with average $-22.8\pm0.5$ and then determined the redshift compatible with the observed I-band magnitude taking into account the evolution, K-correction and Galactic absorption.
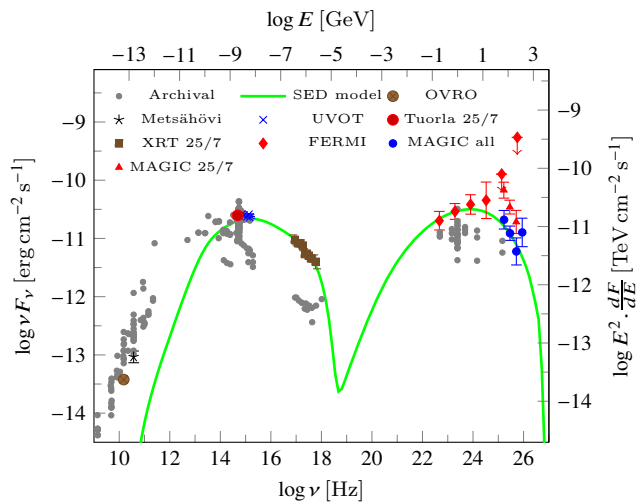
\begin{figure}
\begin{tikzpicture}
\pgfplotsset{
             compat = newest ,
             scaled ticks = false ,
             legend style={at={(0.025,0.975)},legend columns=3,
                           anchor=north west,fill=none,draw=none,
                           font = \tiny},
             every axis plot post/.append style = {error bars/.cd, 
                                                   y dir=both,
                                                   y explicit
                                                   },                           
             every node near coord/.style={font=\tiny},
             width = 0.427\textwidth, %Panels Width.
             %height = 8.9cm ,   %Panels Height.
             select coords between index/.style 2 args={
                  x filter/.code={
                       \ifnum\coordindex<#1\def\pgfmathresult{}\fi
                       \ifnum\coordindex>#2\def\pgfmathresult{}\fi
                                  }
                                 },
             }   
\begin{axis}[name = magic,
             xmin = 9 ,    
             xmax = 27,
             ymin = -14.5,
             ymax = -8.1, 
             ylabel = {$\log  \nu F_{\nu}$\,[erg\,cm$^{-2}$\,s$^{-1}$]},
             xlabel = {$\log  \nu$\,[Hz]},
             xtick pos=left,
             ytick pos=left,
             minor y tick num = 1,
             minor x tick num = 1,
             ytick = {-14,-13,-12,-11,-10,-9},
             xtick = {10,12,14,16,18,20,22,24,26}
             %extra y tick labels = {0.25,0.75, 1.25}
             ]
\addplot[only marks, color=gray, mark=*, 
         mark options={scale=0.5, fill=gray}]
         table [ x = x, y = y, col sep = comma] {sed_arch.csv}; 
\addlegendentry{Archival}
\addplot[green, line width=1pt]	     
          table {sum_blob.dat}; 
\addlegendentry{SED model}
\addplot+[only marks]
         table
         [ select coords between index={0}{0},
           x expr=\thisrowno{0}, 
           y expr=\thisrowno{1},
           y error expr=\thisrowno{2}
         ] {seddata.txt};
\addlegendentry{OVRO}
     
\addplot+[only marks]
         table
         [ select coords between index={1}{1},
           x expr=\thisrowno{0}, 
           y expr=\thisrowno{1},
           y error expr=\thisrowno{2}
         ] {seddata.txt};
\addlegendentry{Mets{\"a}h{\"o}vi}

\addplot+[only marks,mark=x]
         table
         [ select coords between index={3}{8},
           x expr=\thisrowno{0}, 
           y expr=\thisrowno{1},
           y error expr=\thisrowno{2}
         ] {seddata.txt};
\addlegendentry{UVOT} 

\addplot+[only marks]
         table
         [ select coords between index={2}{2},
           x expr=\thisrowno{0}, 
           y expr=\thisrowno{1},
           y error expr=\thisrowno{2}
         ] {seddata.txt};
\addlegendentry{Tuorla 25/7}

%\addplot+[only marks]
%         table
%         [ select coords between index={45}{45},
%           x expr=\thisrowno{0}, 
%           y expr=\thisrowno{1},
%           y error expr=\thisrowno{2}
%         ] {seddata.txt};
%\addlegendentry{Tuorla-L 57230}
%\addplot+[only marks]
%         table
%         [ select coords between index={31}{39},
%           x expr=\thisrowno{0}, 
%           y expr=\thisrowno{1},
%           y error expr=\thisrowno{2}
%         ] {seddata.txt}; 
%\addlegendentry{XRT-H 23/7}    

\addplot+[only marks, mark options={scale=0.7}]
         table
         [ select coords between index={9}{17},
           x expr=\thisrowno{0}, 
           y expr=\thisrowno{1},
           y error expr=\thisrowno{2}
         ] {seddata.txt}; 
\addlegendentry{XRT 25/7}  

%\addplot[color= blue, only marks, mark=star]
%         table
%         [ select coords between index={40}{44},
%           x expr=\thisrowno{0}, 
%           y expr=\thisrowno{1},
%           y error expr=\thisrowno{2}
%         ] {seddata.txt}; 
%\addlegendentry{XRT-L 27/7}
          
\addplot[color= red, only marks, mark=diamond*]
         table
         [ select coords between index={18}{23},
           x expr=\thisrowno{0}, 
           y expr=\thisrowno{1},
           y error expr=\thisrowno{2}
         ] {seddata.txt};
\addlegendentry{FERMI}  
\draw[->, color=red](axis cs:25.1324510513,-9.8958746099)--(axis cs:25.1324510513,-10.1958746099);  
\draw[->, color=red](axis cs:25.7489459662,-9.2656651864)--(axis cs:25.7489459662,-9.5656651864);        
\addplot[only marks, color= blue, mark options={scale=0.7}]
         table
         [ select coords between index={27}{30},
           x expr=\thisrowno{0}, 
           y expr=\thisrowno{1},
           y error expr=\thisrowno{2}
         ] {seddata.txt};  
\addlegendentry{MAGIC all}  
\addplot[only marks, color =red, mark=triangle*, mark options={scale=0.7}]
         table
         [ select coords between index={24}{26},
           x expr=\thisrowno{0}, 
           y expr=\thisrowno{1},
           y error expr=\thisrowno{2}
         ] {seddata.txt};  
\addlegendentry{MAGIC 25/7}    
\end{axis}

\begin{axis}[minor y tick num = 1,
             minor x tick num = 1,
             xmin = -14.3834509721,    
             xmax = 3.6165490279,
             axis x line*=right,
             xtick = {-13,-11,-9,-7,-5,-3,-1,1,3},
             xlabel={$\log E$\,[GeV]},
             ymin = -14.7041199827,
             ymax = -8.3041199827,
             axis y line*=right,
             ytick = {-14,-13,-12,-11,-10,-9,-8},
             ylabel = {$\log E^2.\frac{dF}{dE}$\,[TeV\,cm$^{-2}$\,s$^{-1}$]}
            ]
\end{axis}
\end{tikzpicture}
\caption{\label{SED}Quasi-simultaneous broad band SED of S2 0109+22 during the MAGIC observations. Archival non-simultaneous data are also shown (grey symbols). See the text for detailed information on the period of observation by different instruments.}
\end{figure}

\subsubsection{Redshift upper limit based on absorption of VHE gamma rays}
The absorption of VHE gamma rays through interaction with the EBL increases with source distance and photon energy. Basic assumptions on the intrinsic spectrum  can be used to infer a limit on the distance of the blazar \citep[e.g.][]{2007ApJ...655L..13M,2010MNRAS.405L..76P}. In order to determine an upper limit for the source distance, we assumed that the intrinsic spectrum is described by a power law or a concave function (i.e. hardness does not increase with energy). The archival data (Fig. \ref{SED}) indicates that the spectrum of the source in the HE gamma-ray band is variable. Considering that the source is not located at $z>1$, we assume the hardest possible spectrum for this redshift as an intrinsic power-law index. As a conservative approach we assume a fixed photon-index limit of 1.5 following \citet{2006Natur.440.1018A} and \citet{2012A&A...542A..59M}. We obtain a 95\% confidence level limit to the S2~0109+22 redshift of $z\leq 0.60$. The value is obtained by means of a maximum likelihood fit to the observed event rates vs. the reconstructed energy, modelling the intrinsic spectrum with a power-law function, using the EBL model of \citet{2011MNRAS.410.2556D}, and performing a scan in redshift. The limit is obtained, following \citet{2001NIMPA.458..745R}, from the resulting profile likelihood vs. redshift, with the intrinsic source parameters, and the background rates vs. reconstructed energy, treated as nuisance parameters. A more conservative limit can be estimated by varying the simulated total light throughput of the instrument by $\pm15\%$. This yields an 95\% upper limit on the redshift of $z\leq 0.67$. To estimate the uncertainties caused by EBL model selection, we test eight different EBL models \citep[i.e.][]{2008A&A...487..837F,2010A&A...515A..19K,2010ApJ...712..238F,2012MNRAS.422.3189G,2012ApJ...758L..13H,2013ApJ...768..197I,2016ApJ...827....6S}. The results show that the uncertainties due to EBL model selection are negligible compared with the instrumental uncertainties. Finally, in order to verify the assumed intrinsic photon index (1.5), we compare the results with the ones obtained by assuming the photon index in Section \ref{sec3-2} ($\Gamma=1.81 \pm 0.14$). The comparison shows that results are consistent with each other.

The estimated redshift ($z=0.36\pm0.07$) and the calculated redshift 95\% upper limit ($z\leq 0.67$) in this paper are consistent with the value reported by \citet[][$z>0.35$]{2016MNRAS.458.2836P}. Therefore, we used $z=0.35$ based on the accuracy of the technique and other uncertainties, to calculate the intrinsic properties of the source.

\begin{figure}
\includegraphics[width=0.47\textwidth]{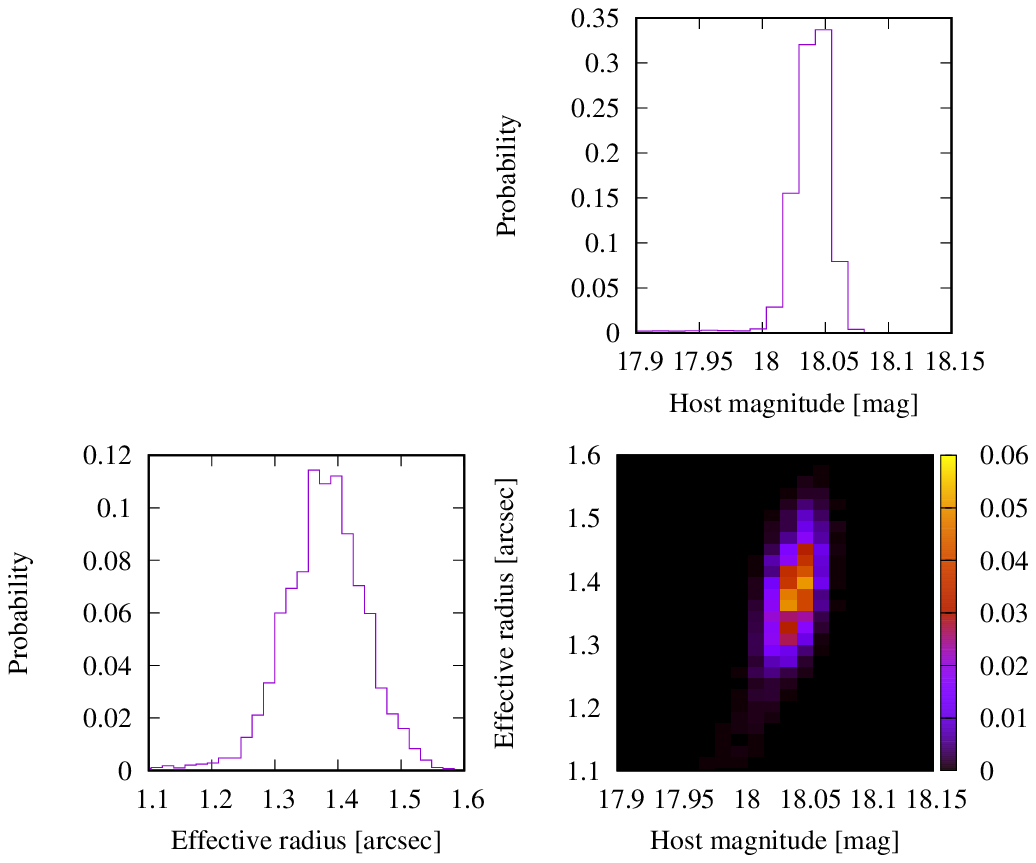}
\caption{\label{hostplot}Marginalized posterior distributions of the host galaxy flux and effective radius \textit{(bottom-right)}. The likelihood distribution of effective radius \textit{(bottom-left)} and host galaxy flux \textit{(top-right)}. The color is proportional to the probability.}
\end{figure}

\subsection{Spectral energy distribution}{\label{3-6}}
In this section, we first present the physical modelling of the SED based on the quasi-simultaneous data described in previous sections. Then, we use a mathematical approach in order to calculate the location of SED peaks in other epochs using archival data (as well as the quasi-simultaneous data near the flare night).

\subsubsection{Broad band SED}
\label{BBSED}
In Figure \ref{SED}, we plot the broad-band SED of S2~0109+22 using the multi-frequency data described in previous sections. For the SED modelling, the HE and VHE gamma-ray spectra are constructed from MAGIC and \textit{Fermi}-LAT data obtained on MJD 57225-57231. The VHE gamma-ray data are corrected for the EBL absorption effect using the \citet{2011MNRAS.410.2556D} model. The VHE gamma-ray spectrum is dominated by the signal from the flare night. However, for the night of the flare, we do not have enough strictly simultaneous data to produce a robust model. As discussed in Section \ref{sec3-1}, the low significance of the signal outside the flare night did not allow us to construct a low-state SED. For X-ray, UV and optical, we selected the data points which are near the flare night, to avoid averaging a variable source with different distribution of observation times during the MAGIC campaign. The \textit{Swift}-UVOT and \textit{Swift}-XRT data are used to reproduce the UV and X-ray spectra of the source on MJD 57228.41. The optical data point, obtained by the KVA telescope on MJD 57228.22, is corrected for Galactic extinction. The host galaxy contribution to the  optical flux is neglected (see Sect. \ref{sec3-5-1}). The radio data points were collected on MJD 57227.41 and 57227.08 in 15 and 37\,GHz respectively, but are not used for SED modelling (see below). 

The quasi-simultaneous SED was modelled using a one-zone synchrotron self-Compton model \citep{2003ApJ...593..667M}. It assumes a spherical, relativistically moving emission region characterized by its radius $R$, magnetic field B and Doppler factor $\delta$. It contains an electron population following a broken power-law distribution with index $p_1$ for $\gamma_\text{min}<\gamma<\gamma_\text{break}$ and $p_2$ for $\gamma_\text{break}<\gamma<\gamma_\text{max}$. The normalization of this electron distribution at $\gamma$=1 is $K$. We use the redshift of $z=0.35$ for the source (see Sect. \ref{sec3-5}).

The goodness of the fitted model is judged by a $\chi^2$-test ($\chi^2$/d.o.f = 22.4/16) assuming fixed $\gamma_\text{min}=1.0 \times 10^{3}$ because there is no instrument available to probe the energy range where the influence of $\gamma_\text{min}$ would be significant. Therefore, the curve represents only one possible set of SED parameters. The other parameters used for the model are: $R=5.5 \times 10^{16}$ cm, $B=0.054$~G, $\delta=21.7$, $\gamma_\text{break}=1.2 \times 10^{4}$, $\gamma_\text{max}=4.5 \times 10^{5}$, $p_1=1.94$, $p_2=3.68$ and $K=3.1 \times 10^3$ cm$^{-3}$. The assumed emission region size is compatible with a daily variability time scale. There is no evidence of a shorter variability time scale in the multi wavelength data during the MAGIC campaign. The parameters are rather typical for TeV BL Lac objects \citep[See e.g.][]{2010MNRAS.401.1570T}.

The one-zone model does not reproduce the spectrum at the lowest frequencies, since the emission is self-absorbed below the millimetre band. It is generally assumed that this emission is produced in the outer regions of the jet. This is in agreement with the results in Section \ref{3-4-1}, where no connection between the long-term behaviour of the optical and radio bands was found for this source. Moreover, the location of the SED peaks are roughly estimated to be $\log \nu_{\text{sync}}\simeq 15.4$ and $\log \nu_{\text{IC}}\simeq 23.3$.

\subsubsection{SED peaks}
\label{syncpeak}
In order to determine the peak frequencies of the SED components, we fitted simultaneously two log-parabolic spectra \citep[e.g][]{2004A&A...413..489M}, one for the synchrotron peak and another for the Inverse Compton (IC), to the SED of the source. We try to calculate the location of the SED peaks for two different states. First, we extracted the archival data from the ASI Space Science Data Center\footnote{\url{http://www.asdc.asi.it/}}. Since the archival data are non-simultaneous and $\nu_{\text{peak}}$ is known to change with the activity state in blazars \citep[e.g.][]{2009ApJ...705.1624A}, we can expect the fitted $\nu_{\text{peak}}$ to depend on the frequencies covered and on the number of observing epochs. To roughly estimate how much this could affect $\nu_{\text{peak}}$ we constructed 4 different samples from the archival data, one representing a high state, another for a low state and two `mixed' states. The archival data indicate that the source is an intermediate synchrotron peak BL Lac object based on the classifications in \citet{2010ApJ...716...30A} with $\log \nu_{\text{sync}}=14.4 \pm 0.1$ and $\log \nu_{\text{IC}}=22.8 \pm 0.2$, which is consistent with the source classification reported by \citet{1999ApJ...525..127L, 2000A&A...361..480D, 2001MNRAS.325.1109B, 2004MNRAS.348.1379C}.

In the second step we used the quasi-simultaneous data described in section \ref{BBSED}. The locations of the peaks are $\log \nu_{\text{sync}}=15.1 \pm 0.5$ and $\log \nu_{\text{IC}}=23.1 \pm 0.2$, which are consistent with the results obtained from the physical modelling described in Section \ref{BBSED}. Table \ref{synchtab} shows the summary of the SED peaks using different approaches and datasets. Based on the broad-band SED modelled for this dataset, the X-ray emission is purely synchrotron, which is normal for HSP BL Lac objects \citep[e.g.][]{2010ApJ...708L.100A}. The historical X-ray observations of 2006 (Table \ref{tab_xray}) show a hard X-ray spectral index ($\Gamma_X=2.06\pm 0.05$) which is in good agreement with the broad-band SED reported by \citet{2004MNRAS.348.1379C} and the normal case for LSP and ISP BL Lac objects \citep[e.g.][]{2017ApJ...844...58P}. Therefore, there is a hint of a transition from intermediate to high synchrotron peak during the MAGIC observation period. The transition is not only in the peak but the whole SED is appearing as a typical X-ray bright HSP SED \citep[e.g. PKS 2155-304:][and references therein]{2012A&A...539A.149H}. 

\section{Summary}
\label{sec4}
S2~0109+22 was discovered for the first time in the HE gamma-ray band by the \textit{Fermi}-LAT during the first three months of sky-survey operation in 2008 \citep{2009ApJ...700..597A}. Previous EGRET upper limits are reported in \citet{2004MNRAS.348.1379C}.

In this paper, we reported the first VHE gamma-ray detection of S2~0109+22 by MAGIC in 2015 July. The MAGIC observation was triggered by the source high state in HE gamma rays. During the MAGIC observation campaign, the HE gamma-ray LC does not show variability on a daily time scale, while the constant fit to VHE gamma-ray flux was rejected with $3\,\sigma$ level of confidence.

We performed a long-term and a short-term multi-frequency study of the source, from radio to VHE gamma rays and compare the source to other TeV blazars. The summary of the main outcomes are: 
\begin{enumerate}
\item Compared to the sample of 21 known variable TeV BL Lac objects (Fig. \ref{histo2}), the observed VHE gamma-ray flux from S2~0109+22 is relatively low. The predicted low state VHE gamma-ray flux by \citet{2017A&A...608A..68F} is below the sensitivity of the current generation of IACTs. Moreover, the source was not detected after its flaring activity by MAGIC. Therefore, this source will be a good candidate to be monitored by the future Cherenkov Telescope Array (CTA) in order to characterize its VHE gamma-ray temporal behaviour and its connection to lower energy bands.
\item The brighter-harder trend is clear in the X-ray band (Table \ref{tab_xray} and Fig. \ref{spec_flux}). Similar behaviour was observed for many TeV BL Lac objects \citep{2017ApJ...841..123P}. However, such a trend is not present in the VHE gamma-ray (Table \ref{tab_tev_spec} and Fig. \ref{vhe_sed}) and HE gamma-ray (Fig. \ref{fig_lc}, panel b and c) bands, but this could be due to large error bars in these bands. The absence of such a correlation in HE and VHE gamma-ray bands for non-HSP BL Lac objects is widely discussed in the context of the `blazar sequence' \citep[see][and references therein]{2015ApJ...810...14A}.
\item In the long-term optical and radio LC (Fig. \ref{fig_longterm}), there was no correlation peak between 15\,GHz and optical flux. This fact suggests that, unlike for many other TeV blazars, the optical and radio emission do not originate from the same region or that the correlation is too complex to be probed by the method found in \citet{2016A&A...593A..98L}.
\item The optical fractional polarization and polarization angle of S2~0109+22 are more variable than found for typical high-energy BL Lac objects \citep[e.g.][]{2016A&A...596A..78H}.
\item We used two methods to estimate the redshift of the source. The result of the photometric host-galaxy method is $z = 0.36\pm0.07$, whereas the 95\% upper limit estimation based on the absorption of VHE gamma-ray emission, assuming the EBL model described in \citet{2011MNRAS.410.2556D}, gives $z \leq 0.67$. The estimated redshifts are in agreement with the one derived by \citet{2016MNRAS.458.2836P}.
\item When comparing the quasi-simultaneous SED presented in this paper with archival data obtained from the ASI Space Science Data Center, there is a hint of intermediate to high synchrotron peak transition. This has been previously suggested for PKS~0301-243 \citep{2013A&A...559A.136H} and 1ES~1011+496 in 2008 \citep{2016MNRAS.459.2286A}.
\item The broad-band SED of S2~0109+22 (Sect. \ref{BBSED}) reveals that the parameters of a single-zone SSC model are rather typical for TeV BL Lac objects. Comparing the SED parameters with the ones reported in \citet{2004MNRAS.348.1379C} reveals that the magnetic field strength is an order of magnitude weaker. Weaker magnetic field energy density ($U_B=B^2/8\pi$) increases the radiation to magnetic energy ratio ($U_{\text{rad}}/U_{B}=L_{\text{IC}}/L_{\text{sync}}$). Therefore, SSC luminosity component increases to the level above the sensitivity of VHE gamma-ray instruments.

\end{enumerate}
The long-term radio to optical and optical polarization behaviour of the source agree with the classification of the source as an ISP BL Lac object, which are still a minority in the class of TeV blazars. However, there is a hint of type transition as discussed in Section \ref{syncpeak} based on the multi epoch comparison of the SED peak locations and X-ray behaviour of the source. In order to precisely characterize the source-type transition behaviour, more simultaneous multi-wavelength observations during different flux states are needed. Such observations can be performed when CTA enables us to detect VHE gamma-ray emission also during the low state of the source. Moreover, considering the increased SSC luminosity, high polarization degree in the optical and high X-ray luminosity of the source make this source an ideal candidate for physical modelling when the X-ray and soft-gamma-ray (MeV) polarization observations become available by instruments such as Imaging X-ray Polarimetry Explorer \citep[IXPE,][]{2016SPIE.9905E..17W}, e-ASTROGAM \citep{2017ExA....44...25D}, and All-sky Medium Energy Gamma-ray Observatory \citep[AMEGO,][]{2017HEAD...1610313M}.

%%%%%%%%%%
\section*{Acknowledgments}
%%%%%%%%%%
Part of this work is based on archival data, software or online services provided by the Space Science Data Center - ASI.

%
% MAGIC STANDARD ACKNOWLEDGEMENTS
% LAST UPDATE: 13 September 2017
% Author: M. Doro
%
We would like to thank the Instituto de Astrof\'{\i}sica de Canarias for the excellent working conditions at the Observatorio del Roque de los Muchachos in La Palma. The financial support of the German BMBF and MPG, the Italian INFN and INAF, the Swiss National Fund SNF, the ERDF under the Spanish MINECO (FPA2015-69818-P, FPA2012-36668, FPA2015-68378-P, FPA2015-69210-C6-2-R, FPA2015-69210-C6-4-R, FPA2015-69210-C6-6-R, AYA2015-71042-P, AYA2016-76012-C3-1-P, ESP2015-71662-C2-2-P, CSD2009-00064), and the Japanese JSPS and MEXT is gratefully acknowledged. This work was also supported by the Spanish Centro de Excelencia ``Severo Ochoa'' SEV-2012-0234 and SEV-2015-0548, and Unidad de Excelencia ``Mar\'{\i}a de Maeztu'' MDM-2014-0369, by the Croatian Science Foundation (HrZZ) Project IP-2016-06-9782 and the University of Rijeka Project 13.12.1.3.02, by the DFG Collaborative Research Centers SFB823/C4 and SFB876/C3, the Polish National Research Centre grant UMO-2016/22/M/ST9/00382 and by the Brazilian MCTIC, CNPq and FAPERJ.

The \textit{Fermi}-LAT Collaboration acknowledges generous ongoing support from a number of agencies and institutes that have supported both the development and the operation of the LAT as well as scientific data analysis. These include the National Aeronautics and Space Administration and the Department of Energy in the United States, the Commissariat \`a l'Energie Atomique and the Centre National de la Recherche Scientifique / Institut National de Physique Nucl\'eaire et de Physique des Particules in France, the Agenzia Spaziale Italiana and the Istituto Nazionale di Fisica Nucleare in Italy, the Ministry of Education, Culture, Sports, Science and Technology (MEXT), High Energy Accelerator Research Organization (KEK) and Japan Aerospace Exploration Agency (JAXA) in Japan, and the K.~A.~Wallenberg Foundation, the Swedish Research Council and the Swedish National Space Board in Sweden. Additional support for science analysis during the operations phase is gratefully acknowledged from the Istituto Nazionale di Astrofisica in Italy and the Centre National d'\'Etudes Spatiales in France. This work performed in part under DOE Contract DEAC02-76SF00515.

The OVRO 40-m monitoring program is supported in part by NASA grants NNX08AW31G, NNX11A043G and NNX14AQ89G, and NSF grants AST-0808050 and AST-1109911.

%%%%%%%%%%%%%%%%%%%%%%%%%%%%%%%%%%%%%%%%%%%%%%%%%%



\bibliographystyle{mnras}
\bibliography{ref.bib} % if your bibtex file is called example.bib

\section*{affiliations}
$^{1}$ {Universit\`a di Udine, and INFN Trieste, I-33100 Udine, Italy} \\
$^{2}$ {National Institute for Astrophysics (INAF), I-00136 Rome, Italy} \\
$^{3}$ {Universit\`a di Padova and INFN, I-35131 Padova, Italy} \\
$^{4}$ {Technische Universit\"at Dortmund, D-44221 Dortmund, Germany} \\
$^{5}$ {Croatian MAGIC Consortium: University of Rijeka, 51000 Rijeka, University of Split - FESB, 21000 Split,  University of Zagreb - FER, 10000 Zagreb, University of Osijek, 31000 Osijek and Rudjer Boskovic Institute, 10000 Zagreb, Croatia.} \\
$^{6}$ {Saha Institute of Nuclear Physics, HBNI, 1/AF Bidhannagar, Salt Lake, Sector-1, Kolkata 700064, India} \\
$^{7}$ {Max-Planck-Institut f\"ur Physik, D-80805 M\"unchen, Germany} \\
$^{8}$ {now at Centro Brasileiro de Pesquisas F\'isicas (CBPF), 22290-180 URCA, Rio de Janeiro (RJ), Brasil} \\
$^{9}$ {Unidad de Part\'iculas y Cosmolog\'ia (UPARCOS), Universidad Complutense, E-28040 Madrid, Spain} \\
$^{10}$ {Inst. de Astrof\'isica de Canarias, E-38200 La Laguna, and Universidad de La Laguna, Dpto. Astrof\'isica, E-38206 La Laguna, Tenerife, Spain} \\
$^{11}$ {University of \L\'od\'z, Department of Astrophysics, PL-90236 \L\'od\'z, Poland} \\
$^{12}$ {Deutsches Elektronen-Synchrotron (DESY), D-15738 Zeuthen, Germany} \\
$^{13}$ {ETH Zurich, CH-8093 Zurich, Switzerland} \\
$^{14}$ {Institut de F\'isica d'Altes Energies (IFAE), The Barcelona Institute of Science and Technology (BIST), E-08193 Bellaterra (Barcelona), Spain} \\
$^{15}$ {Universit\`a  di Siena and INFN Pisa, I-53100 Siena, Italy} \\
$^{16}$ {Universit\`a di Pisa, and INFN Pisa, I-56126 Pisa, Italy} \\
$^{17}$ {Universit\"at W\"urzburg, D-97074 W\"urzburg, Germany} \\
$^{18}$ {Finnish MAGIC Consortium: Tuorla Observatory and Finnish Centre of Astronomy with ESO (FINCA), University of Turku, Vaisalantie 20, FI-21500 Piikki\"o, Astronomy Division, University of Oulu, FIN-90014 University of Oulu, Finland} \\
$^{19}$ {Departament de F\'isica, and CERES-IEEC, Universitat Aut\'onoma de Barcelona, E-08193 Bellaterra, Spain} \\
$^{20}$ {Japanese MAGIC Consortium: ICRR, The University of Tokyo, 277-8582 Chiba, Japan; Department of Physics, Kyoto University, 606-8502 Kyoto, Japan; Tokai University, 259-1292 Kanagawa, Japan; RIKEN, 351-0198 Saitama, Japan} \\
$^{21}$ {Inst. for Nucl. Research and Nucl. Energy, Bulgarian Academy of Sciences, BG-1784 Sofia, Bulgaria} \\
$^{22}$ {Universitat de Barcelona, ICC, IEEC-UB, E-08028 Barcelona, Spain} \\
$^{23}$ {Humboldt University of Berlin, Institut f\"ur Physik D-12489 Berlin Germany}\\
$^{24}$ {also at Dipartimento di Fisica, Universit\`a di Trieste, I-34127 Trieste, Italy}\\
$^{25}$ {also at Port d'Informaci\'o Cient\'ifica (PIC) E-08193 Bellaterra (Barcelona) Spain}\\
$^{26}$ {also at INAF-Trieste and Dept. of Physics \& Astronomy, University of Bologna},\\
$^{27}$ {Agenzia Spaziale Italiana (ASI) Space Science Data Center, I-00133 Roma, Italy)},\\
$^{28}$ {Istituto Nazionale di Fisica Nucleare (INFN), Sezione di Perugia, I-06123 Perugia, Italy},\\
$^{29}$ {Departamento de Astronom\'ia, Universidad de Chile, Camino El Observatorio 1515, Las Condes, Santiago, Chile},\\
$^{30}$ {Tuorla Observatory, University of Turku, Väisäläntie 20, FI-21500 Piikkiö, Finland},\\
$^{31}$ {Aalto University Mets\"ahovi Radio Observatory, Mets\"ahovintie 114, 02540 Kylm\"al\"a, Finland},\\
$^{32}$ {Aalto University Department of Electronics and Nanoengineering,
P.O. BOX 15500, FI-00076 AALTO, Finland},\\
$^{33}$ {Tartu Observatory, Observatooriumi 1, 61602 T\~{o}ravere, Estonia},\\
$^{34}$ {Owens Valley Radio Observatory, California Institute of Technology, Pasadena, CA 91125, USA},

\begin{table*}
\caption{\label{tab_tev_flux}The VHE gamma-ray flux of S2~0109+22}              % title of Table
\centering   
\renewcommand{\baselinestretch}{1.4}\large\normalsize                                 
\begin{tabular}{cccc}          
\hline\hline 
 
\multirow{2}{*}{MJD}&Integration time&$F_{>100\,\text{GeV}}\times 10^{-11}$	   &\multirow{2}{*}{Notes}	\\ 
		    &[s]	     &[\,ph\,cm$^{-2}$\,s$^{-1}$]&		\\ 
\hline  
57225.15   & 4462 &$3.0  \pm 1.3$& \\
57226.15   & 4175 &$4.2  \pm 1.3$& \\
57227.15   & 4609 &$3.6  \pm 1.3$& \\
57228.15   & 5049 &$9.3  \pm 1.4$& Highest observed flux\\
57229.15   & 5249 &$3.8  \pm 1.2$& \\
57230.15   & 4234 &2.0  & 95\% C.L. Upper-limit\\
57231.15  & 5580 &2.3  & 95\% C.L. Upper-limit\\
\hline
\end{tabular}
\end{table*}

\begin{table*}
\caption{\label{tab_tev_spec}The VHE gamma-ray spectrum parameters of S2~0109+22}              % title of Table
\centering   
\renewcommand{\baselinestretch}{1.4}\large\normalsize                                 
\begin{tabular}{lccccc}          
\hline\hline 
 
\multirow{2}{*}{Data set}&$F_0\times 10^{-10}$				&\multirow{2}{*}{$\Gamma$}	&$E_{\text{dec}}$	 &\multirow{2}{*}{$\chi^2/\text{d.o.f.}$}	&Fit Probability	\\ 
			 &[TeV\,cm$^{-2}$\,s$^{-1}$]&				&[GeV]	  	 &                      		&[\%]			\\ 
\hline  
25 Jul 2015 (observed)	&$	11.7\pm1.3	$&$	3.69\pm0.20	$&$	119.43	$&	0.56/3	&	91	\\
25 Jul 2015 (intrinsic)	&$	15.6\pm1.9	$&$	3.07\pm0.30	$&$	119.43	$&	4.92/5	&	43	\\
\hline

All data (observed)	&$	2.5\pm0.3	$&$	3.45\pm0.22	$&$	137.13	$&	1.43	/	2	&	49	\\
All data (intrinsic)	&$	4.2\pm0.5	$&$	2.92\pm0.32	$&$	130.95	$&	9.05	/	7	&	25	\\
\hline
\end{tabular}
\end{table*}

\begin{table*}
\caption{\label{tab_xray}The X-ray properties of S2~0109+22}              
\centering   
\renewcommand{\baselinestretch}{1.4}\large\normalsize                                 
                    
\begin{tabular}{ccccccc}          
\hline\hline                       
\multirow{2}{*}{MJD}	&Exposure time&$F(2-10\,\text{keV})$	&$F(0.3-10\,\text{keV})$	&\multirow{2}{*}{$\Gamma_X$}	 &\multirow{2}{*}{$\chi^2_{\text{reduced}}/\text{d.o.f.}$}	&\multirow{2}{*}{Observation ID}	\\ 
\cline{3-4}
	&[s]&\multicolumn{2}{c}{$\times 10^{-12}$\,[erg\,cm$^{-2}$\,s$^{-1}$]}	&		&	  &\\ 
\hline  
$53762.93\pm0.07	$&1993 &$	0.32\pm0.14	$&$	1.25\pm0.34	$&$	2.69\pm0.23	$&	1.24	/	2	&	00035001001	\\
$53887.45\pm0.44	$&17998 &$	1.01\pm0.07	$&$	2.05\pm0.17	$&$	2.06\pm0.05	$&	1.17	/	49	&	00035001003	\\
$57224.99\pm0.04	$&3951 &$	1.40\pm0.13	$&$	5.26\pm0.44	$&$	2.66\pm0.06	$&	1.05	/	31	&	00040849003	\\
$57225.99\pm0.04	$&3961 &$	5.34\pm0.32	$&$	15.88\pm0.90$&$	2.46\pm0.04	$&	1.20	/	73	&	00040849004	\\
$57226.47\pm0.44	$&3316 &$	2.29\pm0.21	$&$	8.26\pm0.60	$&$	2.63\pm0.06	$&	1.32	/	39	&	00040849005	\\
$57228.45\pm0.04	$&2939 &$	3.80\pm0.29	$&$	12.90\pm0.75$&$	2.58\pm0.05	$&	0.96	/	50	&	00040849006	\\
$57229.39\pm0.31	$&2968 &$	1.10\pm0.16	$&$	4.01\pm0.43	$&$	2.63\pm0.09	$&	0.68	/	16	&	00040849007	\\
$57230.36\pm0.34	$&2038 &$	0.39\pm0.11	$&$	1.41\pm0.31	$&$	2.64\pm0.18	$&	0.17	/	3	&	00040849008	\\
$57231.59\pm0.04	$&1516 &$	0.83\pm0.44	$&$	1.81\pm0.99	$&$	2.14\pm0.32	$&	0.05	/	1	&	00040849010	\\
$57235.87\pm0.01	$&1411 &$	2.94\pm0.44	$&$	8.92\pm1.21	$&$	2.48\pm0.10	$&	0.88	/	14	&	00040849011	\\
\hline
\end{tabular}
\end{table*}

\begin{table*}
\caption{\label{synchtab}Location of SED peaks calculated based on different approaches and states described in Section \ref{3-6}}              
\centering   
\renewcommand{\baselinestretch}{1.4}\large\normalsize                                 
                    
\begin{tabular}{ccccc}          
\hline\hline                       
Dataset &Method &State &$\log \nu_{\text{synch}}$	&$\log \nu_{\text{IC}}$	\\ 
\hline  
\multirow{4}{*}{Archival}&\multirow{4}{*}{Mathematical}& Low & 14.4&22.9\\
&&High &14.6 & 22.9\\
&&Mixed 1 &14.3&22.7\\
&&Mixed 2 &14.5&23.1\\
\hline
Quasi-Simultaneous&Mathematical&-- &15.1&23.1		\\
\hline
Quasi-Simultaneous&Physical Modelling& -- &15.4&23.3 \\
\hline
\end{tabular}
\end{table*}

%%%%%%%%%%%%%%%%%%%%%%%%%%%%%%%%%%%%%%%%%%%%%%%%%%

%%%%%%%%%%%%%%%%% APPENDICES %%%%%%%%%%%%%%%%%%%%%

\appendix

%%%%%%%%%%%%%%%%%%%%%%%%%%%%%%%%%%%%%%%%%%%%%%%%%%

\bsp	% typesetting comment
\label{lastpage}
\end{document}